\documentclass[a4paper,11pt,preprint]{article}
\pdfoutput=1

\usepackage{jcappub}
\usepackage{multirow}
\usepackage{graphicx}
\usepackage{dcolumn}
\usepackage{bm}
\usepackage{color}
\usepackage{epsfig}
\usepackage{hyperref}
\usepackage{natbib}
\usepackage{float}
\usepackage{array}
\usepackage{multirow}
\usepackage{ragged2e}
\usepackage{adjustbox}
\usepackage{dsfont}

\newcommand{\overbar}[1]{\mkern 1.5mu\overline{\mkern-1.5mu#1\mkern-1.5mu}\mkern 1.5mu}

\makeatletter
\newcommand\xleftrightarrow[2][]{%
	\ext@arrow 9999{\longleftrightarrowfill@}{#1}{#2}}
\newcommand\longleftrightarrowfill@{%
	\arrowfill@\leftarrow\relbar\rightarrow}
\makeatother

\title{Ability of LIGO and LISA to probe the equation of state of the early Universe}

\newcommand{\addressEPFL}{Institute of Physics, Laboratory of Particle Physics and Cosmology (LPPC), \'Ecole Polytechnique F\'ed\'erale de Lausanne (EPFL), CH-1015 Lausanne, Switzerland.}
\author{Daniel G. Figueroa}
\author{and Erwin H. Tanin}
\affiliation{\addressEPFL}

\emailAdd{daniel.figueroa@epfl.ch}
\emailAdd{ehtanin@gmail.com}
\date{\today}

\abstract{The expansion history of the Universe between the end of inflation and the onset of radiation-domination (RD) is currently unknown. If the equation of state during this period is stiffer than that of radiation, $w > 1/3$, the gravitational wave (GW) background from inflation acquires a blue-tilt ${d\log\rho_{\rm GW}\over d\log f} = {2(w-1/3)\over (w+1/3)} > 0$ at frequencies $f \gg f_{\rm RD}$ corresponding to modes re-entering the horizon during the stiff-domination (SD), where $f_{\rm RD}$ is the frequency today of the horizon scale at the SD-to-RD transition. We characterized in detail the transfer function of the GW energy density spectrum, considering both 'instant' and smooth modelings of the SD-to-RD transition. The shape of the spectrum is controlled by $w$, $f_{\rm RD}$, and $H_{\rm inf}$ (the Hubble scale of inflation). We determined the parameter space compatible with a detection of this signal by LIGO and LISA, including possible changes in the number of relativistic degrees of freedom, and the presence of a tensor tilt. Consistency with upper bounds on stochastic GW backgrounds, however, rules out a significant fraction of the observable parameter space. We find that this renders the signal unobservable by Advanced LIGO, in all cases. The GW background remains detectable by LISA, though only in a small island of parameter space, corresponding to scenarios with an equation of state in the range $0.46 \lesssim w \lesssim 0.56$ and a high inflationary scale $H_{\rm inf} \gtrsim 10^{13}$ GeV, but low reheating temperature $1~{\rm MeV} \lesssim T_{\rm RD} \lesssim 150$ MeV (equivalently, $10^{-11}~{\rm Hz} \lesssim f_{\rm RD} \lesssim 3.6\cdot10^{-9}~{\rm Hz}$). Implications for early Universe scenarios resting upon an SD epoch are briefly discussed.}

\begin{document}
\maketitle

\section{Introduction}

Observations of the cosmic microwave background (CMB) and the large scale structures of the Universe strongly support the idea that the Universe underwent a period of cosmic inflation at its early stages. Apart from solving the horizon and flatness problems, and providing the appropriate initial conditions for primordial density perturbations, inflation is a very natural phenomenon. The most recent measurement of the B-mode polarization anisotropies of the CMB~\cite{Akrami:2018odb,Ade:2018gkx} sets a bound on the inflationary Hubble rate which corresponds to an energy scale $\lesssim 10^{16}\text{ GeV}$. At some point below this energy scale, the energy budget of the universe must be converted into a thermal bath of radiation in order to switch to the standard hot Big Bang cosmology, from which point the expansion history is well known. To explain the observed abundance of light elements in the universe, Big Bang Nucleosynthesis (BBN) must take place during the radiation domination (RD) epoch, starting at a temperature of around $T_{\rm BBN}\sim 1 \text{ MeV}$, when the energy budget is dominated by photons and relativistic neutrinos. For consistency, the RD epoch must begin therefore before the onset of BBN. With the period of inflation ending at an energy scale $\lesssim 10^{16}\text{ GeV}$ and the onset of RD occurring at least above $\sim 1 \text{ MeV}$, we are left with an unknown intermediary period which may span up to $\sim 19$ orders of magnitude in energy scale.

During inflation, tensor metric perturbations generated from quantum fluctuations are spatially stretched to scales exponentially larger than the inflationary Hubble radius. This process results in a (quasi-)scale invariant tensor power spectrum at superhorizon scales~\cite{Grishchuk:1974ny,Starobinsky:1979ty, Rubakov:1982df,Fabbri:1983us}, with a very small red tilt. After inflation, the horizon begins to grow faster than the redshifting of length scales, and the tensor modes re-enter the horizon successively as the Hubble radius catches up with each wavelength. Once a tensor mode crosses inside the horizon, it becomes part of the stochastic background of gravitational waves (GWs). Different tensor modes re-enter the horizon at different times, and hence propagate through different periods of the evolution of the Universe after they have become sub-horizon.  For tensor modes crossing during RD,  the resulting present GW energy density spectrum is (quasi-)scale invariant, with the same tilt as the original super-horizon inflationary tensor power spectrum. For periods of expansion when the energy budget of the Universe is not dominated by relativistic species, the (quasi-)scale invariance is broken, and the resulting GW spectrum becomes significantly tilted in the frequency range corresponding to the modes crossing the horizon during such period(s). The expansion history of the Universe is imprinted in this way in the power spectrum of the freely-lingering primordial GW background that we can observe today.
In other words, we might detect or constrain the post-inflation expansion history by attempting to detect the relic GWs of inflationary origin, see e.g.~\cite{Giovannini:1998bp,Giovannini:1999bh,Boyle:2005se, Watanabe:2006qe, Boyle:2007zx,Kuroyanagi:2008ye,Kuroyanagi:2010mm,Kuroyanagi:2014qza}. In turn, the expansion history will inform us about the matter fields driving the expansion.

To characterize the post-inflation pre-BBN expansion history, we consider an intermediate epoch taking place between the end of inflation and the onset of RD, with an EoS parameter $w \neq 1/3$. In scenarios where the inflaton oscillates around the minimum of its potential after inflation, an effective EoS (averaged over inflaton oscillations) emerges~\cite{Turner:1983he,Lozanov:2016hid}, lying within the range $0 < w < 1/3$ for $V(\phi) \propto \phi^p$, $p < 4$. There is however $a~priori$ no reason, neither theoretical nor observational, to exclude a {\it stiff} case {\small$1/3 < w \leq 1$}.  A period with a stiff EoS can actually be achieved naturally in a universe dominated by the kinetic energy of a scalar field after inflation, either through inflaton oscillations~\cite{Turner:1983he} under a steep potential (e.g. $V(\phi) \propto \phi^p$ with $p > 4$), or simply by an abrupt drop of the potential for large values of the inflaton. In the latter case it is particularly easy and natural to obtain an EoS close to unity $w\simeq 1$. We note that $w \leq 1$ emerges as a natural upper bound from the requirement that the sound speed $c_{\text{s}}^2=\partial p/\partial \rho \equiv w \leq 1$ of a fluid does not exceed the speed of light. Nevertheless, we expect that in general field theory constructions $w$ can approach unity only from below, so we will consider only $w < 1$. 
Scenarios with a stiff dominated (SD) epoch between inflation and RD are particularly appealing from observational point of view: the stiff period induces a blue-tilt in the GW energy spectrum at large frequencies~\cite{Giovannini:1998bp,Giovannini:1999bh,Riazuelo:2000fc,Sahni:2001qp,Tashiro:2003qp,Boyle:2007zx,Giovannini:2008zg,Giovannini:2008tm,Cui:2017ufi,Caprini:2018mtu,Cui:2018rwi}, opening up the possibility of detection of this GW background by upcoming GW direct-detection experiments. Furthermore, the possibility of a stiff epoch is also well motivated on the theory side, by various early Universe scenarios for which the implementation of a post-inflationary stiff period is crucial. For instance, in {\it Quintessential inflation}~\cite{Peebles:1998qn,Peloso:1999dm,Huey:2001ae,Majumdar:2001mm,Dimopoulos:2001ix,Wetterich:2013jsa,Wetterich:2014gaa,Hossain:2014xha,Rubio:2017gty} the inflationary epoch is followed by a period where the universe is dominated by the kinetic energy of the inflaton, with the potential adjusted to describe the observed dark energy as a quintessence field. In the original {\it gravitational reheating} formulation~\cite{Ford:1986sy,Spokoiny:1993kt} the Universe is reheated by the decay products of inflationary spectator fields if the Universe undergoes a sufficiently long stiff epoch after inflation. Even though basic implementations of such gravitational reheating scenarios have been shown to be inconsistent with BBN/CMB constraints~\cite{Figueroa:2018twl}, unnatural $ad~hoc$ constructions are still viable. Furthermore, variants that can naturally avoid the inconsistency have been also proposed, like the {\it Higgs-reheating} scenario~\cite{Figueroa:2016dsc}, where the Standard Model (SM) Higgs is a spectator field with a non-minimal coupling to gravity and the Universe is reheated into SM relativistic species (decay products of the Higgs), if inflation is followed by an SD period. The same mechanism can actually be realized with generic self-interacting scalar fields~\cite{Dimopoulos:2018wfg}, other than the SM Higgs. See~\cite{Opferkuch:2019zbd} for a recent re-analysis of the idea.

Models with blue-tilted inflationary GW spectrum due to the presence of an SD epoch have been studied in various contexts \cite{Giovannini:1998bp, Giovannini:1999bh, Giovannini:2008zg, Giovannini:2009kg, Tashiro:2003qp, Sahni:2001qp, Boyle:2005se, Boyle:2007zx, Li:2016mmc}.  The resulting GW energy spectrum can be characterized by three quantities: the Hubble rate $H_{\text{inf}}$ during inflation, the equation state parameter $w$ during the stiff epoch, and the Hubble/energy scale at the SD-to-RD transition, parametrized by the redshifted frequency today $f_{\rm RD}$ corresponding to such scale. In general, a blue-tilted GW background can be probed with a variety of experiments, see e.g.~\cite{Zhao:2011bg,Zhao:2013bba,Kuroyanagi:2014nba,Jinno:2014qka,Lentati:2015qwp,Lasky:2015lej,Arzoumanian:2015liz,Liu:2015psa,Kuroyanagi:2018csn,DEramo:2019tit,Bernal:2019lpc}. The aim of our present work is to assess the ability of the {\it Advanced Laser Interferometric Gravitational wave Observatory} (aLIGO), as well as of  the (will-be) first generation space-based GW detector, the {\it Laser Interferometric Space Antenna} (LISA), to probe the parameter space spanned by $H_{\rm inf}$, $w$, and $f_{\rm RD}$.  We assume inflation is well described by a quasi-{\it de Sitter} stage, so that the inflationary tensor tilt is expected to be only slightly-red tilted, according to the standard consistency condition [c.f.~Eq.~(\ref{eq:consistencyCondition})]. A study in this spirit was initiated in~\cite{Boyle:2005se,Boyle:2007zx}, but with the SD-to-RD energy scale fixed to its smallest possible value, $T_{\rm BBN}\sim 10^{-3}\text{ GeV}$, so that $f_{\rm RD}$ was fixed to its lowest possible value $f_{\rm BBN} \sim 10^{-11}$ Hz. In our present work we present a systematic exploration of the observability of the full parameter space $\lbrace w, H_{\rm inf}, f_{\rm RD}\rbrace$. 

As a consistent cosmological history must preserve the success of BBN, we need to prevent the presence of a stiff epoch from changing significantly the expansion rate during BBN. There are two ways by which the latter can happen. First, if the stiff epoch does not end in time before the start of BBN the expansion rate would certainly deviate significantly from that of the $\Lambda$CDM. Second, even if RD starts in time, GWs should not carry too much energy, or otherwise the expansion rate during BBN (or CMB decoupling for this matter) would still be sufficiently altered. Thus, in order to be consistent with BBN/CMB, bounds must be put on the parameters $\lbrace w, H_{\rm inf}, f_{\rm RD}\rbrace$, as these control both the duration of the stiff period and the shape of the GW spectrum (and hence the energy carried by the GWs). As we will see, these constraints will restrict severely the ability of detectors to observe the blue tilted GW background due to an SD epoch. We find that the above constraints render the signal completely inaccessible to the observational window of aLIGO, independently of the parameter space. Whilst the background remains detectable by LISA, it is only observable if the inflationary scale is as large as $H_{\rm inf} \gtrsim 10^{13}$ GeV (corresponding to at least $\mathcal{O}(10)\%$ of its current upper bound), $f_{\rm RD}$ lies in the range $10^{-11}~{\rm Hz} \lesssim f_{\rm RD} \lesssim 3.6\cdot10^{-9}~{\rm Hz}$, or equivalently the Universe becomes RD at sufficiently low temperatures $1~{\rm MeV} \lesssim T_{\rm RD} \lesssim 150$ MeV (i.e.~the SD spoch spans many decades in energy scale), and the stiff EoS is confined within the narrow range $0.46 \lesssim w \lesssim 0.56$. This corresponds to a small island in the parameter space .

The paper is divided as follows. In Section~\ref{sec:Inflation}, we briefly review the form of the spectrum of GWs from inflation. In Section~\ref{sec:GWspectrum}, we derive the energy density spectrum of the inflationary GW background in the presence of an SD epoch, considering two possible modelings for the SD-to-RD transition: an 'instantaneous' transition with a sharp jump in the EoS, and a 'smooth' transition modeled by the evolution of two fluid components, one made of radiation and another by a scalar field dominated by its own kinetic energy. In Section~\ref{sec:Results}, we first quantify the full parameter space $\lbrace w, H_{\rm inf}, f_{\rm RD}\rbrace$ that can be probed by LISA and aLIGO. We then restrict such parameter space to be consistent with upper bounds on stochastic GW backgrounds from BBN and CMB considerations. We also extend our analysis to include possible changes in the number of relativistic degrees of freedom ($dof$), and the presence of a small red tensor tilt, as motivated in slow-roll inflation. In Section~\ref{sec:Discussion}, we summarize our findings and discuss briefly the implications for some early Universe scenarios resting upon the presence of an SD epoch.
 
From now on, {\small$m_p = {1/\sqrt{8\pi G}} \simeq 2.44\cdot10^{18}$ GeV} is the reduced Planck mass, {\small$a(\tau)$} is the scale factor, {\small$\tau \equiv \int {dt\over a(t)}$} is the conformal time, and we use the Friedman-Lema\^itre-Robertson-Walker (FLRW) metric $ds^2 = a^2(\tau)\eta_{\mu\nu}dx^\mu dx^\nu$ as the background metric. 

\section{Gravitational waves}
\label{sec:Inflation}

A tensor-perturbed FLRW metric can be written as
\begin{align}
ds^2=a^2(\tau) \left[d\tau^2-\left(\delta_{ij}+h_{ij}\right)dx^idx^j\right]\,, \label{perturbedmetric}
\end{align}
where we assume the perturbations to be transverse and traceless $\partial_j h_{ij} = h^i_i = 0$, so that they can be identified with gravitational waves (GWs). Expanding the Einstein equations to linear order $O(h)$ gives
\begin{align}
h_{ij}^{\prime\prime}+2\frac{a^{\prime}}{a}h_{ij}^{\prime}-\nabla^2 h_{ij} = 0\,,\label{eomspatial}
\end{align}
where primes denote derivatives with respect to the conformal time $\tau$. To bring \eqref{eomspatial} to a more useful form, we perform a spatial Fourier- and polarization-mode decomposition
\begin{align}
h_{ij}(\tau,\mathbf{x})&=\sum_{\lambda}\int \frac{d^3\textbf{k}}{(2\pi)^3} e^{i\mathbf{k}.\mathbf{x}} \epsilon_{ij}^\lambda(\mathbf{k})h^\lambda_{\mathbf{k}}(\tau)\,,
\end{align}
where $\lambda = +, \times$ stands for the polarization states and $\epsilon_{ij}^\lambda(\mathbf{k})$ are a basis of polarization tensors satisfying $\epsilon_{ij}^\lambda(\mathbf{k})=\epsilon_{ji}^\lambda(\mathbf{k})$, $\epsilon_{ii}^\lambda(\mathbf{k})=0$, $k_i\epsilon_{ij}^\lambda(\mathbf{k})=0$, $\epsilon_{ij}^\lambda(\mathbf{k})=\epsilon_{ij}^{\lambda^*}(-\mathbf{k})$, and $\epsilon_{ij}^\lambda(\mathbf{k})\epsilon_{ij}^{\sigma^*}(\mathbf{k})=2\delta^{\lambda\sigma}$. These bring us to the GW equation of motion
\begin{align}
h_{k}^{\prime\prime}+2\frac{a^\prime}{a}h_{k}^\prime+k^2h_{k}=0\,, \label{eom}
\end{align}
where we have suppressed the polarization indices $\lambda$, as we assume that the GW spectrum is unpolarized $\left<|h_{\mathbf{k}}^{+}|\right>=\left<|h_{\mathbf{k}}^{\times}|\right> \equiv \left<|h_{\mathbf{k}}|\right>$. Assuming that the background metric is isotropic, we also write $h_k = h_{\mathbf{k}}$, with $k\equiv|\mathbf{k}|$. 

A useful quantity to characterize a GW background is the tensor power spectrum $\Delta_h(\tau, k)$, defined through the following relation
\begin{align}
\left<h_{ij}(\tau,\mathbf{x})h^{ij}(\tau,\mathbf{x})\right>&\equiv\int \frac{dk}{k} \Delta_h^2(\tau,k) ~~~~ \Longleftrightarrow ~~~~  
\Delta_h^2(\tau,k)=\frac{2k^3}{\pi^2} \left<|h_k(\tau)|^2\right>\,,
\label{deltah}
\end{align}
where $\left<...\right>$ denotes a statistical ensemble average.

\subsection{Inflationary spectrum}

In this paper we focus on the tensor modes generated during inflation as they were spatially stretched past the inflationary Hubble radius. At the end of inflation, these modes represent superhorizon tensor perturbations with an almost scale invariant power spectrum~\cite{Caprini:2018mtu}
\begin{align}
\Delta_{h,\text{inf}}^2(k)\simeq \frac{2}{\pi^2}\left(\frac{H_{\text{inf}}}{ m_{\text{p}}}\right)^2\left({k\over k_p}\right)^{n_t} \label{inf}\,,
\end{align}
with $n_t$ a spectral tilt, $k_{p}$ a pivot scale, 
and $H_{\rm inf}$ the Hubble rate when the mode $k_p$ exited the horizon during inflation. The presence of the tilt stems from the fact that the inflationary phase cannot be perfectly {\it de Sitter}. Nevertheless, the spectrum is expected to be only slightly red-tilted in slow-roll inflation, with the spectral index 'slow-roll suppressed' as
\begin{align}\label{eq:consistencyCondition}
n_t \simeq -2\epsilon \simeq -{r_p\over8} \,,
\end{align}
where $r_{p}$ 
is the tensor-to-scalar ratio evaluated at the scale $k_p$, constrained by the most recent analysis of the B-mode polarization anisotropies of the CMB at a scale $k_p/a_0 = 0.002~{\rm Mpc}^{-1}$, 
as $r_p \leq 0.064$~\cite{Akrami:2018odb,Ade:2018gkx}. This bound actually implies an upper bound on the energy scale of inflation, which must be constrained as
\begin{eqnarray}\label{eq:Hmax}
H_{\rm inf} \lesssim H_{\rm max} \simeq 6.6\cdot 10^{13}\,{\rm GeV}\,.
\end{eqnarray}
Furthermore, the upper bound on $r_p$ also implies that the red-tilted spectral index can only be very small $-n_t \leq 0.008 \ll 1$. The tensor spectrum is therefore very close to be exactly scale-invariant, at least at around the CMB scales. Actually, in the absence of running of the spectral index, the amplitude of the tensor spectrum would fall only by a factor $\sim (10^{26})^{-0.008} \sim 0.6$ during the $\log(e^{60}) \sim $ 26 orders of magnitude separating the CMB scales and the Hubble radius at the end of inflation. Therefore, for simplicity, we will consider from now on an exact scale-invariant inflationary spectrum, as this gives an excellent approximation. We will comment on deviations from this assumption in Sect.~\ref{s:spectraltilt}. 

From a theoretical perspective, it is convenient to work with the power spectrum $\Delta_h^2(k)$, as  it is precisely this quantity that is predicted by inflation to be approximately scale invariant. During the evolution of the Universe after inflation, when the tensor modes cross inside the Hubble radius, they become a stochastic background of gravitational waves (GWs). In order to quantify the ability of GW direct detection experiments to measure the inflationary GW background, it is customary to express the amount of GWs in terms of their energy density spectrum  (at sub-horizon scales) $\Omega_{\text{GW}}$, defined as the GW energy density $\rho_{\text{GW}}$ per unit logarithmic comoving wavenumber interval, normalized to the critical density $\rho_{\text{crit}}=3m_p^2H^2$ \cite{Caprini:2018mtu},
\begin{align}
\Omega_{\text{GW}}(\tau, k) &\equiv\dfrac{1}{\rho_{\text{crit}}}\dfrac{d\rho_{\text{GW}}(\tau,k)}{d\ln k} = \frac{k^2}{12a^2(\tau)H^2(\tau)}\Delta_h^2(\tau,k)\,,  \label{energyspectrum}
\end{align}
It is customary to factorize the tensor power spectrum at arbitrary times as a function of the primordial inflationary spectrum $\Delta_{h,\text{inf}}^2(k)$ [c.f.~Eq.~(\ref{inf})] by means of a transfer function~\cite{Boyle:2005se}
\begin{align}
\Delta_h^2(\tau,k)\equiv T_h(\tau,k) \Delta_{h,\text{inf}}^2(k)\,,~~~~ T_h(\tau,k) \equiv {1\over 2}\left(a_k\over a(\tau)\right)^2\,,\label{transferfunc}
\end{align}
which characterizes the expansion history between the moment of horizon re-entry $\tau = \tau_k$ of a given mode $k$, defined as $a_kH_k \equiv k$ with $a_k \equiv a(\tau_k)$, $H_k\equiv H(\tau_k)$, and a later moment $\tau > \tau_k$. For the power spectrum today we will use the notation $T_h(k) \equiv T_h(\tau_0,k)$. Note that the factor $1\over2$ in Eq.~(\ref{transferfunc}) is simply due to averaging over harmonic oscillations of the modes deep inside the horizon.

If we assume that immediately after inflation, the Universe became radiation dominated (RD) with equation of state $w = 1/3$, the resulting present-day GW energy density spectrum is (quasi-)scale invariant, for the frequency range corresponding to the modes crossing the Hubble radius during RD. Setting $n_t = 0$ and averaging over oscillations, the amplitude of the $plateau$ characterizing the energy density spectrum today is
\begin{eqnarray}\label{eq:InfGWtodayRD}
 \Omega_{\rm GW}^{(0)}{\Big |}_{\rm plateau}  &\simeq& ~~\mathcal{G}_k { \Omega_{\rm rad}^{(0)}\over12\pi^2}\left(H_{\rm inf}\over m_p\right)^2  \simeq ~1\cdot10^{-16}\left(H_{\rm inf}\over H_{\rm max}\right)^2\,,
\end{eqnarray}
where we have used $k = a_kH_k$ and introduced the RD transfer function~\cite{Boyle:2005se}
\begin{eqnarray}\label{eq:TransferAndGk}
 T_h(k) \simeq {1\over2}\left({a_k\over a_0}\right)^2 \simeq {1\over2}\mathcal{G}_k\Omega_{\rm rad}^{(0)}\left({a_0H_0\over a_k H_k}\right)^2\,,~~~
\mathcal{G}_k \equiv \left(g_{*,k}\over g_{*,0}\right)\left(g_{s,0}\over g_{s,k}\right)^{4/3}\,.
\end{eqnarray}
In the $rhs$ of Eq.~(\ref{eq:InfGWtodayRD}) we have used $\Omega_{\rm rad}^{(0)} \simeq 9\cdot 10^{-5}$, $g_{s,0} \simeq 3.91$, $g_{*,0} = 3.36$, and $g_{s,k} \simeq g_{*,k} \simeq 106.75$ (so that $\mathcal{G}_k \simeq 0.39$). For simplicity we have considered $g_{s,k}, g_{*,k}$ equal to the Standard Model (SM) degrees of freedom ($dof$) before the electroweak symmetry breaking, and hence independent of $k$. In reality the number of SM relativistic $dof$ change with the scale, but we postpone the discussion of this spectral distortion to Section~\ref{s:changesindof}. For the time being we simply consider an identical suppression for all the modes as $\mathcal{G}_k \sim 0.39$.

Eq.~(\ref{eq:InfGWtodayRD}) describes the amplitude of the $plateau$ of the inflationary GW (quasi-)scale invariant  energy density spectrum today, corresponding to the modes that crossed the horizon during RD. However, if after inflation there is a transient period of evolution with EoS $w \neq 1/3$, before RD is established, the resulting GW energy density spectrum today will no longer remain scale-invariant. As we will see next, the spectrum today will actually consist of two parts: a high-frequency branch, corresponding to the modes that crossed the horizon during the transient epoch, and a (quasi-)scale invariant branch corresponding to the modes that crossed the horizon during RD\footnote{There is yet another part of the spectrum, corresponding to modes that crossed the Hubble radius after matter-radiation equality, which behaves as $\Omega^0_{\rm GW} \propto 1/k^2$. This corresponds to very small frequencies today $f \lesssim 10^{-16}$ Hz, and hence we will not be concerned with such low frequency end of the spectrum, as it only affects the CMB and it cannot be probed by direct-detection GW experiments.}.

 \section{Inflationary spectrum in the presence of a stiff epoch}
 \label{sec:GWspectrum}

Let us consider now that there is a period in the early Universe, spanning from the end of inflation till the onset of RD, with EoS $w \neq 1/3$ (possibly depending on time). In standard single field slow-roll scenarios the inflaton exhibits a minimum in the potential around which it oscillates in the period following inflation. For an inflaton potential of the form $V \propto \phi^{2n}$, an effective (oscillation averaged) EoS emerges as $\overbar w \simeq (n-1)/(n+1)$~\cite{Turner:1983he}. For $n = 2$  ($V \propto \phi^{4}$) we obtain a RD period with $\overbar w \simeq 1/3$, while for $n = 1$ ($V \propto \phi^2$) we obtain instead a matter dominated (MD) era with $\overbar w \simeq 0$. Interestingly, for $n \geq 3$ we obtain a \textit{stiff dominated} (SD) period with EoS $\overbar w > 1/3$.

In general the effective EoS in the epoch following immediately after inflation must fall in the range {$-1/3 < \overbar w < 1$}. Even though it is common to assume that {$0 \leq \bar w \leq 1/3$}, there is $a~priori$ no reason (theoretical or observational) to exclude the {\it stiff} case {$1/3 < \overbar w < 1$}. In this paper we are particularly interested in exploring this latter possibility. In fact, a post-inflationary period with a stiff EoS can be realized easily in a generic model of inflation. For example, in scalar singlet driven inflation, the slow-roll condition is achieved by simply demanding {$V \gg K$}, where {$V$} and {$K$} are the inflaton potential and kinetic energy densities. Inflation cannot be sustained however if the potential drops to {$V < K/2$}. Furthermore, if a feature in the inflaton potential allows its value {$V$} to drop much below the kinetic energy {$K$}, the EoS can become stiff after inflation, {$w = (K-V)/(K+V) > 1/3$}. 

A simple realization of an SD regime is obtained by assuming a rapid transition of the potential from {$V \gg K$} during inflation to some small value {$V \ll K$} after inflation. The transition itself would actually trigger the end of inflation, leading to a post-inflationary EoS {$\overbar w \simeq 1 - \mathcal{O}(V/K)$}. In general we expect that the EoS can approach unity from below, but never achieve $\overbar w = 1$ exactly, as this would require an exactly flat direction with $V = 0$. A natural scenario where inflation is followed by a KD phase is that of Quintessential-Inflation~\cite{Peebles:1998qn}, where the inflaton potential $V(\phi)$ is engineered so that the necessary transition occurs at the end of inflation, and the potential is also adjusted to describe the observed dark energy as a quintessence field, see e.g.~\cite{Peloso:1999dm,Huey:2001ae,Majumdar:2001mm,Dimopoulos:2001ix,Wetterich:2013jsa,Wetterich:2014gaa,Hossain:2014xha,Rubio:2017gty} for different proposals. As mentioned before, a stiff period can be also engineered through the oscillations of the inflaton with potential $V \propto \phi^{2n}$, $n \geq 3$. 
In this case we note however that the stiff period cannot be sustained for very long, as self-resonant effects lead eventually to a fragmentation of the coherent oscillating inflaton condensate~\cite{Lozanov:2017hjm}.

Considering the presence of an SD period before the onset of RD is actually not only theoretically well motivated but also phenomenologically interesting. As we have mentioned in the Introduction, and as we will show in detail, an SD period with equation of state $\overbar w > 1/3$  induces a large blue-tilt in the high frequency branch of the inflationary GW energy density spectrum~\cite{Giovannini:1998bp}, making this signal possibly observable with direct detection GW experiments. The aim of this paper is to quantify precisely our observational ability to measure such GW background, i.e.~to determine the observable parameter space (we will also present a brief discussion on the implications for particle physics models). For our purposes, the details of the SD model implementation are unimportant. Hence, from now on we will focus on the phenomenology of SD assuming that for some unknown reason there is such a phase following the end of inflation. The background energy density of the inflationary sector evolves after inflation as {$\rho_{\rm tot} =\rho_*\exp\lbrace-3\int {da\over a} (w(a)+ 1)\rbrace$}, with $\rho_* = 3m_p^2H_{*}^2$ the initial energy density at the end of inflation. We will distinguish the Hubble rate at the end of inflation $H_*$ from that during inflation $H_{\rm inf}$ throughout the text, except in Section~\ref{subsec:BBNandCMBbounds} and \ref{subsec:TiltAndDof} where we take $H_{\rm inf}\simeq H_*$ in deriving the numerical results. In general the EoS is determined by the inflaton potential and is a function of time. However we expect it to change only adiabatically during SD, and in any case we can always describe the scaling of the energy density in terms of an effective (logarithmic-averaged) value of the EoS during the stiff period, $3\int {da\over a}(1+w(a)) \equiv 3(1+\overbar w)\log (a/a_*)$, so that {$\rho_{\rm tot} = \rho_*(a/a_*)^{-3(\overbar w+ 1)}$}.

\begin{figure}[t]
	\includegraphics[width=15cm]{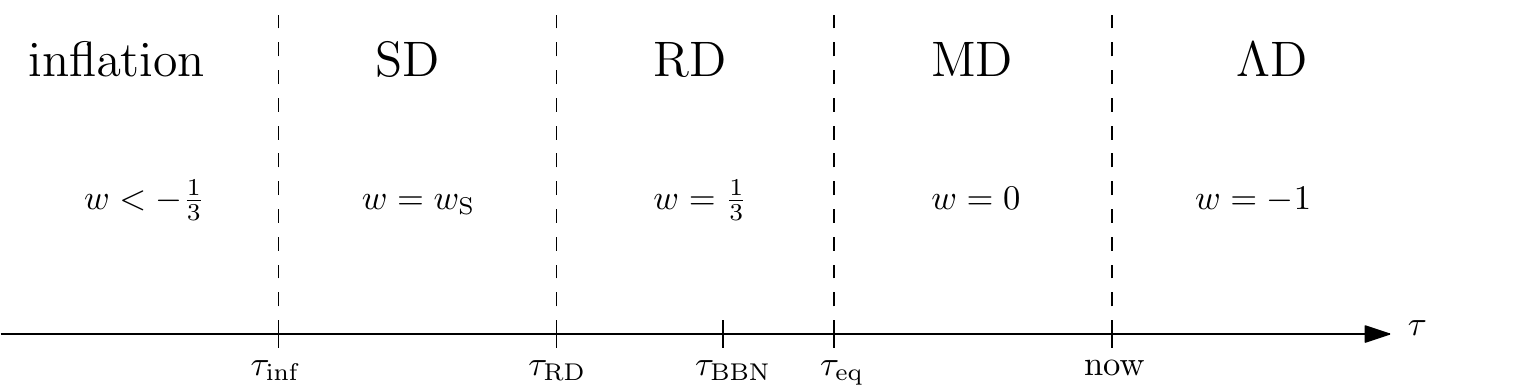}
	\caption{$\Lambda$CDM+inflation expansion history with a stiff epoch.}
	\label{fig:setup}
\end{figure}

Let us consider therefore an SD period between the end of inflation and the onset of RD, with effective EoS $\overbar w \equiv w_{\rm s} > 1/3$ deep inside the SD (i.e.~way before reaching RD). Once the Universe enters into RD, we match the expansion history with that of a Universe with energy budget dominated by the SM radiation $dof$, according to the standard hot Big Bang picture. We sketch the different epochs of the expansion history we consider in Figure~\ref{fig:setup}. From now on, the subscripts $_{\text{*}}$, $_{\text{RD}}$, and $_{\text{BBN}}$, stand for ``evaluation at" or ``corresponding to" the end of inflation/beginning of SD epoch, end of SD epoch/beginning of RD period, and onset of BBN, respectively. In order to not sabotage the success of BBN, a minimal requisite that we need to impose over the assumed expansion history is that the stiff epoch must end before the beginning of BBN, i.e.~$\tau_{\text{RD}}<\tau_{\text{BBN}}$. 

In this section we will solve \eqref{eom} for the aforementioned cosmological scenario, propagate the solution to the present era when it can be detected, and express the result in terms of the GW energy density spectrum today~\eqref{energyspectrum}. Our focus is mainly on the modes that re-enter the horizon during the stiff phase (SD-reentering modes), as they constitute the part of the GW spectrum that is enhanced and hence potentially detectable. The expression for the present-day GW energy spectrum depends on the assumption on how the transition from the SD to RD era takes place. There are two natural cases to consider:

\begin{enumerate}

\item \textbf{Instant transition.} Here the transition from SD to RD is modeled as 'instantaneous', i.e. it occurs in a very short time interval compared to the Hubble timescale at the moment of the transition. The scale factor $a(\tau)$ and Hubble rate $H(\tau)$ are continuous during the transition, but we consider a sudden jump in the effective EoS from $\overbar w = w_{\rm s} > 1/3$ to $w = 1/3$. For example, this could happen if the stiff fluid decays into radiation at some point (which marks the end of the SD epoch) by some process characterized by a timescale much shorter than the instantaneous Hubble time at that moment.
    
\item \textbf{Smooth transition.} Here the SD spoch is driven by a fluid component (typically formed by the inflaton field itself) which dominates the energy budget of the Universe and has a stiff equation of state $w_{\rm s} > 1/3$, but there is also a relatively small amount of radiation present at the end of inflation / onset of SD. The energy densities of the two components scale freely as the Universe expands, i.e.~we assume no interaction between the two sectors. Due to the different scalings, as $\propto a^{-4}$ for radiation and as $\propto a^{-3(1+w_{\rm s})}$ for the stiff fluid, the radiation component eventually dominates the energy budget of the Universe. The transition from SD to RD then occurs smoothly, over a few Hubble times, as the energy density of radiation gradually overtakes that of the stiff fluid.
\end{enumerate}

Since the time-dependence of the scale factor around the SD-to-RD transition is different in the two cases, the GWs that have entered the horizon during SD will evolve differently. In turn, this means that the GW energy spectrum for the modes that re-entered the horizon before and around the SD-to-RD transition will differ in the two cases. The GW energy spectrum can be computed fully analytically in the instant transition case. However, in the smooth transition case we can only compute analytically the asymptotic high and low frequency branches of the GW spectrum, corresponding to the modes that entered the horizon deep inside SD and RD, respectively. From an observational point of view, this is not a problem, as only the high frequency branch of the spectrum can be potentially probed by GW detectors. For completeness, in any case, we will provide a numerical computation  of the full GW spectrum in the smooth transition case.

\subsection{Instant transition}

We consider in this subsection the expansion history shown in Figure~\ref{fig:setup}, assuming the transition from the SD to RD epoch occurs instantaneously, that is, in a time much shorter than the instantaneous Hubble time at the moment of transition. We consider the scale factor and the Hubble rate around the transition as continuous smooth functions, but we model the EoS as a discrete function
\begin{eqnarray}
 \bar w = w_{\rm s}\Theta(\tau_{\rm RD}-\tau) + {1\over3}\Theta(\tau-\tau_{\rm RD})\,,
\end{eqnarray}
where $\Theta(x)$ is the step-function. The energy density of the background is therefore continuous, and scales as 
\begin{eqnarray}
\rho_{\rm tot} = \left\lbrace 
\begin{array}{ll}
 \rho_*(a/a_*)^{-3(1+w_{\rm s})} & ,\, \tau \leq \tau_{\rm RD}~ ({\rm Stiff~Domination})\vspace*{4mm}\\
\rho_{\rm RD}(a/a_{\rm RD})^{-4} & ,\, \tau \geq \tau_{\rm RD} ~({\rm Radiation~Domination})
\end{array}
\right .
\end{eqnarray}
where $\rho_{\rm RD} \equiv \rho_*(a_{\rm RD}/a_*)^{-3(1+w_{\rm s})}$. From here the scale factor during each period can then be solved exactly as
\begin{eqnarray}
a(\tau) &=& a_{*}\left[1+{a_*H_*\over \alpha_{\rm s}}(\tau - \tau_*)\right]^{\alpha_{\rm s}} = a_{*}\left({a_*H_*\over \alpha_{\rm s}}\right)^{\alpha_{\rm s}}[{\tilde\tau}_s(\tau)]^{\alpha_{\rm s}}\,,~~~~ 
\tau_{*}\leq\tau\leq\tau_{\text{RD}} \label{astiff}\\
a(\tau) &=& a_{\rm RD}\left[1+a_{\rm RD}H_{\rm RD}(\tau - \tau_{\rm RD})\right] = a_{\text{RD}}^2H_{\text{RD}}{\tilde\tau}_r(\tau)\,,~~~~ 
\tau_{\text{RD}}\leq \tau \ll\tau_{\text{eq}} \label{arad}    
\end{eqnarray}
and, correspondingly, the Hubble rate $aH(\tau) \equiv {d\log a\over d\tau}$ as
\begin{eqnarray}
aH(\tau) &=& {a_*H_*\over 1+{a_*H_*\over \alpha_{\rm s}}(\tau-\tau_*)} = {\alpha_{\rm s}\over \tilde{\tau}_s(\tau)}\,,~~~~ \tau_{*}\leq\tau\leq\tau_{\text{RD}}\label{eq:Hstiff}\\
aH(\tau) &=& {a_{\rm RD}H_{\rm RD}\over 1+a_{\rm RD}H_{\rm RD}(\tau-\tau_{\rm RD})} =  {1\over \tilde{\tau}_r(\tau)}\,,~~~~ \tau_{\rm RD}\leq\tau\ll \tau_{\text{eq}}\,,\label{eq:Hrad}
\end{eqnarray}
where $\tau_{\text{eq}}$ denotes the time at the radiation-matter equality, and we have defined
\begin{equation}
\alpha_{\rm s} \equiv \frac{2}{1+3w_{\rm s}}\,.
\end{equation}
For convenience, we have introduced the time variables $\tilde{\tau}_{\rm s}$ and $\tilde{\tau}_{\rm r}$ via
\begin{eqnarray}
{a_*H_*\over \alpha_{\rm s}}\tilde{\tau}_s(\tau) \equiv 1+{a_*H_*\over \alpha_{\rm s}}(\tau - \tau_*)\,,~~~~~~
a_{\rm RD}H_{\rm RD}\tilde{\tau}_r(\tau) \equiv 1+a_{\rm RD}H_{\rm RD}(\tau - \tau_{\rm RD})\,. \label{tautilde}
\end{eqnarray} 
From now on, and without loss of generality, we fix $a_* = 1$ and $\tau_* = 0$ at the end of inflation (onset of SD).

During the stiff epoch the GW equation of motion \eqref{eom} reads, using \eqref{eq:Hstiff},
\begin{align}
h_k^{\prime\prime}+\frac{2\alpha_{\rm s}}{\tilde\tau_s}h_k^{\prime}+k^2h_k=0\,, \label{eomconstantw}
\end{align}
where $'$ denotes derivatives with respect to $\tilde\tau_s$ (since $d\tilde{\tau}_s=d\tau$, we do not distinguish between the derivatives with respect to $\tilde{\tau}_s$ or $\tau$). Requiring that the tensor mode function must match the inflationary spectrum $h_k(\tau) = h_k^{\text{inf}}$ and $h_k^\prime(\tau) = 0$ in the superhorizon limit $k\tau \ll 1$, the solution to \eqref{eomconstantw} during the stiff period when $\tau_{*}\leq\tau\leq\tau_{\text{RD}}$ is
\begin{align}
h_k^{\text{(stiff)}}(\tau)=\Gamma\left(\alpha_{\rm s}+\frac{1}{2}\right)\left(\frac{2}{\alpha_{\rm s}y}\right)^{\alpha_{\rm s}-{1\over2}}J_{\alpha_{\rm s}-{1\over2}}(\alpha_{\rm s}y)~h_k^{\text{inf}}\,, ~~~~~~
y \equiv {k\over aH} = {k\tilde\tau_{s}(\tau)\over \alpha_{\rm s}}\,,  \label{stiffsol}
\end{align}
where $J_\nu(x)$ is the Bessel function of the first kind. Using the small argument limit of the Bessel function, $J_{\nu}(x) \simeq x^\nu / (2^\nu \Gamma(\nu+1))$ for $x \ll 1$, we obtain $h_k^{\text{(stiff)}}(\tau) \simeq h_k^{\rm inf}$ when $k \ll  aH$, as it should. The superscript $^{(\text{stiff})}$ indicates that the solution applies in the stiff epoch. We will use analogous notations in what follows.

Using \eqref{eq:Hrad}, the GW equation of motion \eqref{eom} during the the RD epoch reads
\begin{align}
h_k^{\prime\prime}+\frac{2}{\tilde{\tau}_r}h_k^{\prime}+k^2h_k=0\,,\label{eq:eomRD}
\end{align}
where this time $'$ denotes derivatives with respect to $\tilde\tau_r$ (again simply because $d\tilde\tau_r = d\tau$). The solution during the RD era to (\ref{eq:eomRD})  $\tau_{\text{RD}}\leq \tau \ll\tau_{\text{eq}}$, is
\begin{equation}
h_k^{\text{(rad)}}(\tau)=\frac{1}{\sqrt{y}}\left[A(k) J_{1\over2}(y)+B(k) Y_{1\over2}(y)\right]\,, ~~~~~~
y \equiv {k\over aH} = k\tilde\tau_{r}(\tau)\,, \label{radsol}
\end{equation}
where $J_{1/2}(y), Y_{1/2}(y)$ are Bessel functions of the first and second kind and the superscript $^{(\text{rad})}$ indicates that the solution applies during RD. At $\tau=\tau_{\text{RD}}$ this solution must match with solution (\ref{stiffsol}) [and hence simultaneously with $h_k^{\text{inf}}$ if the mode is superhorizon]. Continuity of the tensor modes and their derivatives requires 
\begin{align}
h_k^{\text{(stiff)}}(\tau_{\text{RD}}) = h_k^{\text{(rad)}}(\tau_{\text{RD}})\,,~~~~
h_k^{\prime \text{(stiff)}}(\tau_{\text{RD}}) = h_k^{\prime \text{(rad)}}(\tau_{\text{RD}})
\end{align}
from which we get
\begin{eqnarray}
\left\lbrace \begin{array}{c}
A(k)\vspace*{0.3cm}\\B(k)          
\end{array}
\right\rbrace = {1\over\sqrt{2}}\left(\frac{2}{\alpha_{\rm s}}\right)^{\alpha_{\rm s}}\Gamma\left(\alpha_{\rm s}+\frac{1}{2}\right)\hspace*{-1mm}~\kappa^{1-\alpha}~h_k^{\text{inf}}\times \left\lbrace \begin{array}{c}
a(\kappa)\vspace*{0.3cm}\\b(\kappa)          
\end{array}\right\rbrace\,,
\label{ab}
\end{eqnarray}
where 
\begin{eqnarray}
\kappa \equiv {k\over k_{\rm RD}}\,,
\end{eqnarray}
with $k_{\rm RD} \equiv a_{\rm RD}H_{\rm RD}$, and
\begin{eqnarray}
a(\kappa) =\sqrt{\alpha_{\rm s}}\left(\dfrac{J_{\alpha_{\rm s}-{1\over2}}(\alpha_{\rm s}\kappa)Y_{3\over2}(\kappa)-J_{\alpha_{\rm s}+{1\over2}}(\alpha_{\rm s}\kappa)Y_{1\over2}(\kappa)}{J_{1\over2}(\kappa)Y_{3\over2}(\kappa)-J_{3\over2}(\kappa)Y_{1\over2}(\kappa)}\right)\,,\\
b(\kappa) =\sqrt{\alpha_{\rm s}}\left(\dfrac{J_{\alpha_{\rm s}+{1\over2}}(\alpha_{\rm s}\kappa)J_{1\over2}(\kappa)-J_{\alpha_{\rm s}-{1\over2}}(\alpha_{\rm s}\kappa)J_{3\over2}(\kappa)}{J_{1\over2}(\kappa)Y_{3\over2}(\kappa)-J_{3\over2}(\kappa)Y_{1\over2}(\kappa)}\right)\,. \label{abtilde}
\end{eqnarray}
The sub-horizon $k\tau\gg 1$ and $\tau\gg \tau_{\rm RD}$ limit of \eqref{radsol} is 
\begin{equation}
    h_k^{\text{(rad)}}(\tau)=\sqrt{\frac{2}{\pi y^2}}\,\left[\,A(k)\sin y-B(k)\cos y\,\right]\,,\label{eq:TensorRDsubHubble}
\end{equation}
where we used the large argument expansion of Bessel functions $J_{\nu}(x \gg 1) \simeq \sqrt{2\over\pi x}\sin(x-\delta_\nu)$, $Y_{\nu}(x \gg 1) \simeq \sqrt{2\over\pi x}\cos(x-\delta_\nu)$ and $\tilde{\tau}\approx\tau$. Substituting \eqref{ab} into \eqref{eq:TensorRDsubHubble}, squaring, and averaging over mode oscillations, we obtain
\begin{eqnarray}
\overbar{|h_k^{\rm(rad)}(\tau)|^2} &=&\frac{1}{2\pi y^2}\left(\frac{2}{\alpha_{\rm s}}\right)^{2\alpha_{\rm s}}\Gamma^2\left(\alpha_{\rm s}+\frac{1}{2}\right)\kappa^{2(1-\alpha_{\rm s})}~\mathcal{W}(\kappa)\,|h_k^{\text{inf}}|^2\label{hsquarerad}\\
&=& \left(\frac{a_{\text{RD}}}{a(\tau)}\right)^2\frac{1}{2\pi}\left(\frac{2}{\alpha_{\rm s}}\right)^{2\alpha_{\rm s}}\Gamma^2\left(\alpha_{\rm s}+\frac{1}{2}\right)\kappa^{-2\alpha_{\rm s}}~\mathcal{W}(\kappa)\,|h_k^{\text{inf}}|^2\,,\label{hsquarerad2}
\end{eqnarray}
where we have used $y = \kappa(a/a_{\rm RD})$ in the second line, and defined
\begin{eqnarray}
 \mathcal{W}(\kappa) \equiv a^2(\kappa) + b^2(\kappa) = {\pi \alpha_{\rm s}\over 2 \kappa} \left[ \left(\kappa J_{\alpha_{\rm s}+{1\over2}}(\kappa)-J_{\alpha_{\rm s}-{1\over2}}(\kappa)\right)^2+\kappa^2J^2_{\alpha_{\rm s}-{1\over2}}(\kappa)\right]\,.\label{eq:WindowFunct}
\end{eqnarray}
It can be shown that subhorizon modes always scale as $h_k\propto a^{-1}$ regardless of how the scale factor $a$ evolves with time. Thus, even though the second expression of $\overbar{|h_k^{\rm(rad)}(\tau)|^2}$ c.f.~Eq.~(\ref{hsquarerad2}), was derived in the RD epoch, once its time evolution (i.e.~the damping of the tensors due to the expansion of the Universe) is written as $\propto a^{-2}(\tau)$ the expression remains valid for the subsequent epochs as well. This is, however, only true for modes that became subhorizon before the moment of matter-radiation equality at $\tau=\tau_{\rm eq}$, i.e.~for modes $k\gg k_{\rm eq} \equiv a_{\rm eq}H_{\rm eq}$. 

Building the present-day tensor power spectrum $\Delta_{h}^2(\tau_0,k) = {2k^3\over\pi^2}\,\overbar{|h_k^{\rm(rad)}(\tau_0)|^2}$  with \eqref{hsquarerad2} and plugging this into \eqref{energyspectrum}, leads to the present-day energy spectrum for the modes $k\gg k_{\rm eq}$ re-entering the horizon during the SD or RD epochs,
\begin{eqnarray}
\hspace*{-0.5cm}\Omega_{\text{GW}}^{(0)}(k) &\equiv& {k^2 \Delta_{h}^2(\tau_0,k) \over 12 a_0^2 H_0^2} = \frac{a_{\text{RD}}^2k^2}{24\pi a_0^4H_0^2}\left(\frac{2}{\alpha_{\rm s}}\right)^{2\alpha_{\rm s}}\Gamma^2\left(\alpha_{\rm s}+\frac{1}{2}\right)\kappa^{-2\alpha_{\rm s}}~\mathcal{W}(\kappa)\,\Delta_{h,{\rm inf}}^2(k)\nonumber\\ 
   & =&\left(\frac{a_{\text{RD}}}{a_0}\right)^4\left(\frac{H_{\rm RD}}{H_0}\right)^2{1\over12\pi^2}\left(\frac{H_{\text{inf}}}{m_{\text{p}}}\right)^2 {\Gamma^2\left(\alpha_{\rm s}+1/2\right)\over 2^{2(1-\alpha_{\rm s})}\alpha_{\rm s}^{2\alpha_{\rm s}}\Gamma^2({3/2})}\,\mathcal{W}(\kappa)\,\kappa^{2(1-\alpha_{\rm s})}\,,\label{OGW0abrupt}
\end{eqnarray}
where in the last step we have introduced inflationary tensor power spectrum \eqref{inf} (with $n_t = 0$), and used $\kappa \equiv k/k_{\rm RD}$, $k_{\rm RD} = a_{\rm RD}H_{\rm RD}$, and $\pi = 4\Gamma^2(3/2)$. Since here we consider an instant SD-to-RD transition, the radiation energy density is equal to the critical density at the start of RD, $\rho_{\rm rad}(\tau_{\rm RD})=\rho_{\rm crit}(\tau_{\rm RD})=3m_p^2H_{\rm RD}^2$. This and the scaling law of radiation energy density implies\footnote{As we will see later on, an analogous relation in the smooth transition case differs by a factor of 2.}
\begin{equation}
	\left(\frac{a_{\text{RD}}}{a_0}\right)^4\left(\frac{H_{\rm RD}}{H_0}\right)^2=\frac{8\pi G\rho_{\rm rad}(\tau_0)}{3H_0^2} = \Omega_{\rm rad}^{(0)}\left(g_{*,k}\over g_{*,0}\right)\left(g_{s,0}\over g_{s,k}\right)^{4/3}\,.\label{aRDa0abrupt}
\end{equation}
Plugging (\ref{aRDa0abrupt}) into Eq.~(\ref{OGW0abrupt}), using Eq.~(\ref{eq:InfGWtodayRD}) for the inflationary plateau, and expressing the result 
as a function of present-day frequencies $f=k/(2\pi a_0)$, we finally obtain
\begin{eqnarray}\label{eq:GWfullSpectrumInstant}
\Omega_{\text{GW}}^{(0)}(f) = \Omega_{\rm GW}^{(0)}{\Big |}_{\rm plateau} \times \mathcal{W}(f/f_{\rm RD}) \times \mathcal{A}_{\rm s}\,\left({f\over f_{\rm RD}}\right)^{2(1-\alpha_{\rm s})}\,,
\end{eqnarray}
where $f_{\rm RD} \equiv k_{\rm RD}/(2\pi a_0)$ is the frequency corresponding to the horizon scale at the onset of RD, $k_{\rm RD} = a_{\rm RD}H_{\rm RD}$, $\mathcal{W}(x)$ is the window function defined in Eq.~(\ref{eq:WindowFunct}), and we have introduced the constant
\begin{eqnarray}\label{eq:Aconstant}
\mathcal{A}_{\rm s} \equiv {\Gamma^2\left(\alpha_{\rm s}+1/2\right)\over 2^{2(1-\alpha_{\rm s})}\alpha_{\rm s}^{2\alpha_{\rm s}}\Gamma^2({3/2})}\,,
\end{eqnarray}
which ranges as $1  <  \mathcal{A}_{\rm s} < 4/\pi \simeq 1.27$ for $1/3 < w_{\rm s} < 1$.
The {\it window} function $\mathcal{W}(x)$ varies smoothly around the frequencies $f \sim f_{\rm RD}$, and its asymptotic limits at large frequencies $f \gg f_{\rm RD}$ (corresponding to modes crossing the horizon during SD) and small frequencies $f \ll f_{\rm RD}$ (corresponding to modes crossing the horizon during RD), determine the asymptotic behaviour of the energy density spectrum. In particular we obtain
\begin{eqnarray}
 \mathcal{W}(f/f_{\rm RD} \ll 1) \longrightarrow  \mathcal{A}_{\rm s}^{-1}\,\left({f\over f_{\rm RD}}\right)^{-2(1-\alpha_{\rm s})}\,,\hspace*{1cm}
\mathcal{W}(f/f_{\rm RD} \gg 1) \longrightarrow 1 \,,
\end{eqnarray}
and hence
 \begin{eqnarray}\label{eq:GWasympInstant}
\Omega_{\text{GW}}^{(0)}(f) \simeq \Omega_{\rm GW}^{(0)}{\Big |}_{\rm plateau}
\times\left\lbrace
\begin{array}{crl}
1 & \,, & f \ll f_{\rm RD} \vspace*{0.3cm}\\
\mathcal{A}_{\rm s}\,\left({f\over f_{\rm RD}}\right)^{2(1-\alpha_{\rm s})} & \,, &f \gg f_{\rm RD} \\
\end{array}
\right.\,.
\end{eqnarray}
What matters from the point of view of detection prospects of this signal is the fact that the high-frequency branch of the spectrum rises with frequency, exhibiting a significant blue tilt for a stiff EoS $w_{\rm S} > 1/3$,
\begin{eqnarray}
 n_t \equiv {d\log\Omega_{\rm GW}^{(0)}\over d\log f} = 2(1-\alpha_{\rm s}) = 2\left({3 w_{\rm S} -1 \over 3 w_{\rm S}+1}\right)> 0\,,
\end{eqnarray}
which approaches unity $n_t \longrightarrow 1$ as we take $w_{\rm S} \longrightarrow 1$. It is precisely this large tilt that leads us to consider the ability of GW detectors to measure this signal: as we will discuss later, a significant fraction of the parameter space characterizing the shape of the spectrum, $\lbrace w_{\rm S}, f_{\rm RD}, H_{\rm inf}\rbrace $ leads to the high-frequency part of the spectrum being above the sensitivity of LISA and LIGO at their corresponding key frequencies.

The window function characterizes, in a sense, the 'interpolation' around the central frequencies $f \sim f_{\rm RD}$, of the two asymptotic regimes at large $f \gg f_{\rm RD}$ and small $f \ll f_{\rm RD}$ frequencies. In the next section, we will actually compute numerically the exact frequency dependence of the window function $\mathcal{W}(f)$ when the SD-to-RD  transition is not modeled as instantaneous, but rather a smooth transition resulting from the gradual domination of the energy budget in the Universe of an initially small radiation component. From an observational point of view, given the current upper bound on the amplitude of the small frequency $plateau$ [c.f.~Eq.~(\ref{eq:InfGWtodayRD})], the actual frequency dependence of $\mathcal{W}(f)$ around $f \sim f_{\rm RD}$ is irrelevant, as it cannot be observed by direct detection experiments. Nevertheless, as the expansion history of the Universe changes depending on the expansion rate assumed around the SD-to-RD transition, the high frequency branch $f \gg f_{\rm RD}$ of the GW spectrum will experience a slightly different expansion history once the corresponding modes cross inside the horizon. As we will show next, this translates into a correction of the normalization constant $\mathcal{A}_{\rm s}$ characterizing the rising high-frequency branch of the spectrum.

Note that the late-time acceleration correction factor $(\Omega_{\rm m}/\Omega_\Lambda)^2$ pointed out in \cite{Zhang:2005nw} does not apply to our derivation. The correction factor is only needed if we adopt the commonly used fitting formula which can be traced back to Eq.~(16) of~\cite{Turner:1993vb}. This fitting formula is obtained by solving \eqref{eom} numerically in the presence of matter and radiation, but without cosmological constant, and then fitting the resulting transfer function. The fitted transfer function needs to be corrected by the factor $(\Omega_{\rm m}/\Omega_\Lambda)^2$ simply because the cosmological constant was not taken into account in getting the transfer function (see e.g. Section~3 of \cite{Turner:1993vb} or the Appendix of \cite{Kuroyanagi:2009br}). Now, in our derivation all the information about the expansion history between the horizon reentry $\tau_k$ and the present time $\tau_0$, including that of the late cosmological-constant dominance, is contained in the ratio $a_0/a_k$. Thus, it would be incorrect to include the said correction factor. If we set the scale factor today $a_0=1$ and specified a given mode $k$, the correction factor is just the ratio between the scale factor at the moment of horizon crossing $a_k$
for a universe with cosmological constant and that for a universe without cosmological constant. Since we are always considering a complete universe with cosmological 
constant included, there is nothing to correct for in our derivation.

\subsection{Smooth transition}

Let us consider now a situation where at the end of inflation $\tau=\tau_{*}$, the total energy density  $\rho_* \equiv 3m_p^2H_*^2$ is split between a dominant stiff fluid with energy density $\rho_\text{stiff}^* = (1-\epsilon)\rho_*$, $\epsilon \ll 1$, and a subdominant amount of radiation with energy density $\rho_\text{rad}^* = \epsilon \rho_* \ll \rho_\text{stiff}^*$. We assume that the stiff fluid has an equation of state parameter $w_{\rm s} > 1/3$, which we take as constant for simplicity. The expansion of the Universe is then driven by the energy densities of the two fluids, each of which scale freely, as $\rho_{\text{rad}}\propto a^{-4}$ and as $\rho_{\text{stiff}}\propto a^{-3(1+w_\text{S})}$. The outline of the expansion history will roughly follow the sketch shown in Figure\,\ref{fig:setup}, but this time, the transition from SD to RD is slow and smooth, instead of 'instantaneous'. While the stiff fluid dominates, $\rho_{\rm stiff} \gg \rho_{\rm rad}$, the equation of state of the Universe remains approximately constant and equal to $w_{\rm s}$. Since $\rho_{\rm stiff}$ scales down faster than $\rho_{\rm rad}$, there is always a moment, which we denote as $\tau=\tau_{\rm RD}$, at which $\rho_\text{rad}( \tau = \tau_{\rm RD}) = \rho_\text{stiff}(\tau = \tau_{\rm RD})$. This point marks the end of the SD epoch and the beginning of the RD epoch, although the expansion history around this moment is neither purely SD nor RD, but rather dictated by a mixture of the two fluids. In a few Hubble times after $\tau =\tau_{\rm RD}$, the radiation component becomes the energy-dominating fluid, and from then on the expansion history follows that of standard RD embedded in the usual $\Lambda$CDM scenario.

In such smooth SD-to-RD transition, the scale factor cannot be solved analytically, let alone the GW equation of motion. It is, however, possible to work out analytically the blue-tilted high-frequency branch of the GW spectrum corresponding to modes entering the horizon deep inside the SD epoch at $\tau\ll \tau_{\text{RD}}$, when the equation of state of the Universe $w\simeq w_{\rm s}$ is approximately constant. Fortunately, these are the modes that can actually be probed by GW detectors. Far before the SD-to-RD transition takes place, the difference between the previously considered instant transition and the presently considered smooth transition is not yet apparent and the tensor mode function is given by the same expression as in the instant transition case \eqref{stiffsol}. In order to avoid having to deal with the part of the expansion history close to the SD to RD transition where the evolution of the scale factor is not analytically solvable, we employ the trick we used earlier, namely rewriting the tensor power spectrum in terms of the scale factor $\propto a^{-2}$. Once we do that, the resulting expression will be valid in all the subsequent epochs. 

In order to proceed, we need first to obtain the value of $a_{\text{RD}}$ in the two fluid approach. The condition $\rho_\text{rad}(\tau_{\rm RD})=\rho_\text{stiff}(\tau_{\rm RD})$ at $\tau_{\rm RD}$ implies  $\rho_{\text{stiff}}(\tau_{\text{RD}})/\rho_{\text{stiff}}(\tau_{*})=H_{\text{RD}}^2/2H_{*}^2$ and the scaling of the energy density of the stiff fluid gives $
\rho_{\text{stiff}}(\tau_{\text{RD}})/\rho_{\text{stiff}}(\tau_{*})=\left(a_{\text{RD}}/a_{*}\right)^{-3(w_{\rm s}+1)}$. Together, they yield
\begin{equation}
\frac{a_{\text{RD}}}{a_{*}}=\left(2^{1/2}\frac{H_{*}}{H_{\text{RD}}}\right)^{\frac{\alpha_{\rm s}}{1+\alpha_{\rm s}}}\,.\label{aratioHratio}
\end{equation}
Taking the sub-horizon limit of expression (\ref{stiffsol}), squaring it, and averaging over oscillations, we arrive at
\begin{eqnarray}
\overbar{\left|h_{k\gg k_{\rm RD}}^{\text{(stiff)}}(\tau)\right|^2} &=&\frac{1}{2\pi}\Gamma^2\left(\alpha_{\rm s}+\frac{1}{2}\right)\left(\frac{2}{k\tilde\tau_s}\right)^{2\alpha_{\rm s}}\left|h_k^{\text{inf}}\right|^2\,,\label{stiffsubsquare} \nonumber\\
&=& \frac{1}{2\pi}\Gamma^2\left(\alpha_{\rm s}+\frac{1}{2}\right)\left(\frac{2}{\alpha_{\rm s}^2}\right)^{\alpha_{\rm s}} \kappa^{-2\alpha_{\rm s}}\left(\frac{a_{\text{RD}}}{a(\tau)}\right)^{2}\left|h_k^{\text{inf}}\right|^2\,,
\label{stiffsubsquare2} 
\end{eqnarray}
where we recall that $\kappa \equiv k/k_{\rm RD} = f/f_{\rm RD}$,  $a_*H_*\tilde\tau_s(\tau) = {\alpha_{\rm s}} + a_*H_*(\tau-\tau_*)$ [c.f.~(\ref{tautilde})], and in the second line we have used the scale factor $a(\tau) = a_{*}^{1+\alpha_{\rm s}}H_*^{\alpha_{\rm s}}\alpha_{\rm s}^{-\alpha_{\rm s}}[{\tilde\tau}_s(\tau)]^{\alpha_{\rm s}}$ deep inside SD during $\tau_{*} \leq \tau \ll\tau_{\text{RD}}$ [c.f.~(\ref{astiff})], together with (\ref{aratioHratio}) and $k_{\rm RD} \equiv a_{\rm RD}H_{\rm RD}$.

Now that the solution is expressed in terms of the scale factor, it remains valid in all the subsequent epochs, and we can omit the superscript $^{\text{(stiff)}}$. Building the present-day tensor power spectrum $\Delta_{h}^2(\tau_0,k) = {2k^3\over\pi^2}\,\overbar{|h_{k\gg k_{\rm RD}}(\tau_0)|^2}$  with \eqref{hsquarerad2}, and plugging this into \eqref{energyspectrum}, leads to the present-day energy spectrum for the modes $k\gg k_{\rm RD}$ re-entering the horizon during the SD,
\begin{eqnarray}
\Omega_{\text{GW}}^{(0)}(f\gg f_{\rm RD}) &\equiv& {k^2 \Delta_{h}^2(\tau_0,k) \over 12 a_0^2 H_0^2} = \frac{a_{\text{RD}}^2k^2}{24\pi a_0^4H_0^2}\left(\frac{2}{\alpha_{\rm s}^2}\right)^{\alpha_{\rm s}}\Gamma^2\left(\alpha_{\rm s}+\frac{1}{2}\right)\kappa^{-2\alpha_{\rm s}}\,\Delta_{h,{\rm inf}}^2(k)\nonumber\\ 
   & =&\left(\frac{a_{\text{RD}}}{a_0}\right)^4\left(\frac{H_{\rm RD}}{H_0}\right)^2{1\over12\pi^2}\left(\frac{H_{\text{inf}}}{m_{\text{p}}}\right)^2 {\Gamma^2\left(\alpha_{\rm s}+1/2\right)\over 2^{2-\alpha_{\rm s}}\alpha_{\rm s}^{2\alpha_{\rm s}}\Gamma^2({3/2})}\,\kappa^{2(1-\alpha_{\rm s})}\nonumber\\
   & =&  \Omega_{\rm GW}^{(0)}{\Big |}_{\rm plateau} \times \tilde{\mathcal{A}}_{\rm s}\,\left({f\over f_{\rm RD}}\right)^{2(1-\alpha_{\rm s})}\,,~~~~~ \tilde{\mathcal{A}}_{\rm s} \equiv 2^{1-\alpha_{\rm s}}\mathcal{A}_{\rm s}\,,
\label{OGW0}
\end{eqnarray}
where in the second step we have introduced the inflationary tensor power spectrum \eqref{inf} (with $n_t = 0$) and used $\kappa \equiv k/k_{\rm RD}$, $k_{\rm RD} = a_{\rm RD}H_{\rm RD}$ and $\pi = 4\Gamma^2(3/2)$, whereas in the third step we have used that $\kappa = f/f_{\rm RD}$, the definition of $\mathcal{A}_{\rm s}$ [c.f.~Eq.~(\ref{eq:Aconstant})] and of the inflationary $plateau$ [c.f.~(\ref{eq:InfGWtodayRD})], and the fact that in a smooth transition $\rho_{\rm rad}(\tau_{\rm RD})=\rho_{\rm crit}(\tau_{\rm RD})/2=3H_{\rm RD}^2/16\pi G$ which implies
\begin{equation}
\left(\frac{a_{\text{RD}}}{a_0}\right)^4\left(\frac{H_{\rm RD}}{H_0}\right)^2=\frac{2\rho_{\rm rad}(\tau_0)}{3m_p^2H_0^2} = 2\,\Omega_{\rm rad}^{(0)}\left(g_{*,k}\over g_{*,0}\right)\left(g_{s,0}\over g_{s,k}\right)^{4/3}\,.\label{aRDa0smooth}
\end{equation}
We notice that in (\ref{aRDa0smooth}) there is an extra factor of $2$ compared to the analogous expression \eqref{aRDa0abrupt} for the instant transition case. As before, in the final expression of Eq.~(\ref{OGW0}) we absorbed the effects due to the changes in the relativistic $dof$ into $\Omega_{\rm GW}^{(0)}{\Big |}_{\rm plateau}$. 

\begin{figure}[t]
\begin{center}
\includegraphics[width=12cm]{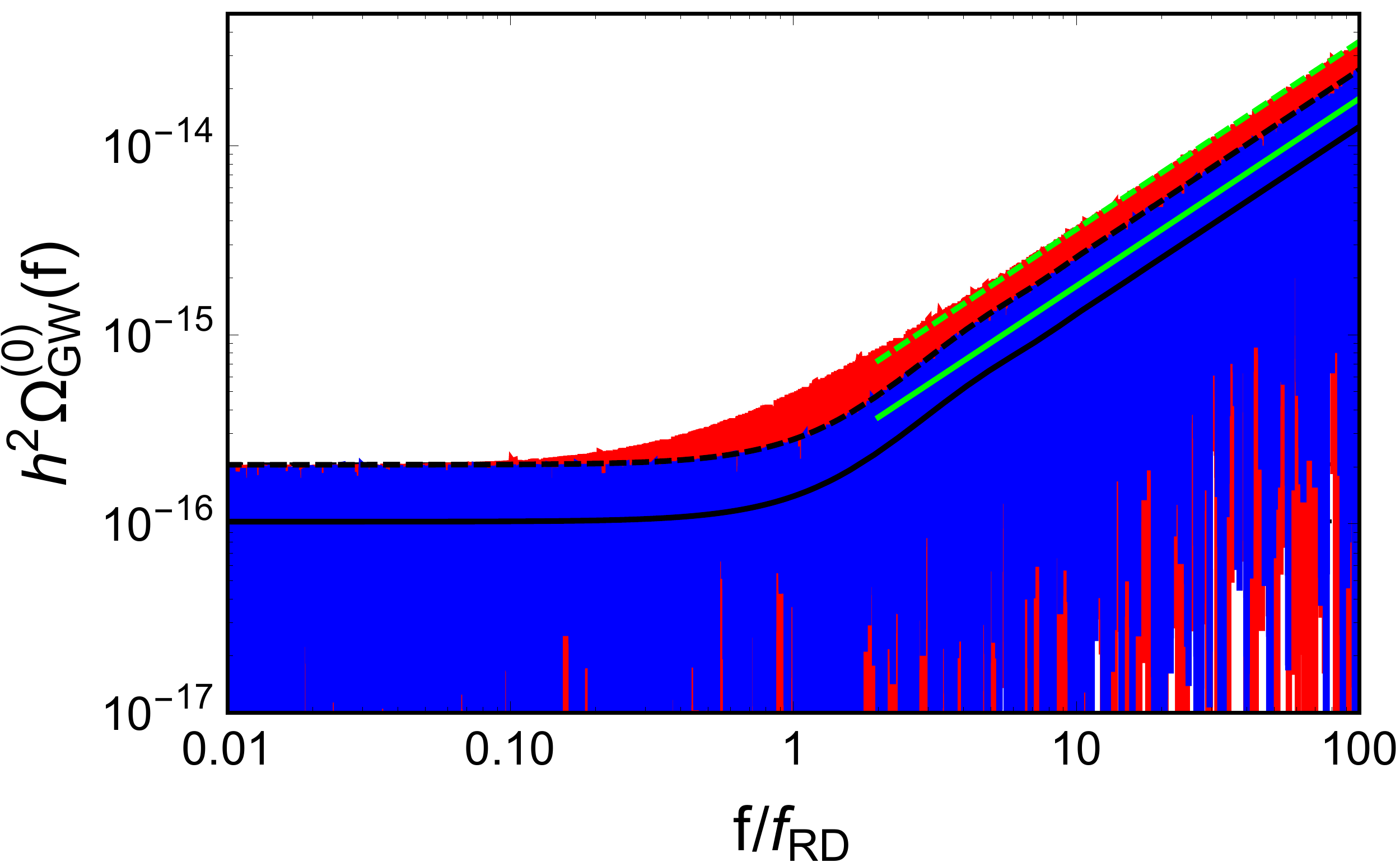}	
\caption{Comparison among different forms of the present-day GW energy density spectra. For the instant transition case we show the full oscillatory solution (blue solid line) computed with Eqs.~(\ref{eq:instantosc})-(\ref{eq:Wosc}), together with its oscillation envelope (black dashed line) and oscillation averaged analytical spectrum (black solid line) computed with Eq.~\eqref{eq:GWfullSpectrumInstant}. For the the smooth transition case we plot the full oscillatory form (red solid line) based on numerical evaluation of Eqs.~(\ref{eq:Fullnum})-(\ref{eq:Numcond2}), and for its high frequency branch the oscillation envelope (green dashed line) and the oscillation averaged analytical spectrum (green solid line) obtained with Eq.~(\ref{OGW0}). For each type of transition the jump between the oscillation envelope (dashed lines) and oscillation average (solid lines) is a factor $2$ exactly, as the tensor mode functions exhibit exact harmonic oscillations at deep inside sub-horizon scales. The difference in amplitude in the high frequency branch between the smooth and the instant transition cases, correspond to a factor $2^{1-\alpha_{\rm s}} \simeq 1.41$, as obtained for $w_{\rm s} = 0.99$.}
\label{fig:GWspectrumNumericAndAnalytic}
\end{center}
\end{figure}

If we compare the expression of the high-frequency branch of the GW energy spectrum we just obtained in the smooth transition case with its instant transition counterpart (\ref{eq:GWasympInstant}), we see that the normalization constant is now a factor $2^{1-\alpha_{\rm s}}$ larger, i.e.~a factor that ranges from 1 (if $w_{\rm s}=1/3$) to $\sqrt{2}$ (if $w_{\rm s}\rightarrow 1$). Therefore, for stiff EoS close to $w_{\rm s} \simeq 1$, the amplitude of the observable high-frequency branch of the spectrum is actually $\sim 40\%$ larger than in the instant transition case. For comparison we plot in Figure~\ref{fig:GWspectrumNumericAndAnalytic} the present GW energy density power spectrum obtained in the instant SD-to-RD transition model, c.f.~Eq.~(\ref{eq:GWfullSpectrumInstant}), together with the high frequency branch obtained in the smooth transition case, c.f.~Eq.~(\ref{OGW0}). For completeness, we also plot the GW spectrum without averaging over oscillations. In the instant transition this corresponds to 
\begin{eqnarray}
\label{eq:instantosc}
 \Omega_{\text{GW}}^{(0)}(f) = \Omega_{\rm GW}^{(0)}{\Big |}_{\rm plateau} \times \mathcal{A}_{\rm s}\,\kappa^{2(1-\alpha_{\rm s})}\times \mathcal{W}_{osc}(\kappa)\,,~~~~ \kappa \equiv {k\over k_{\rm RD}} ={ f\over f_{\rm RD}}\\
 \label{eq:Wosc}
\mathcal{W}_{osc}(\kappa) \equiv  \pi y\left[a(\kappa)J_{1\over2}(y)+b(\kappa)Y_{1\over2}(y)\right]^2\,,~~~~ y \equiv {k\over aH}\,.\hspace*{1cm}
\end{eqnarray}
In a smooth transition we need to obtain the spectrum fully numerically as
\begin{eqnarray}\label{eq:Fullnum}
 \Omega_{\text{GW}}^{(0)}(f) = \Omega_{\rm GW}^{(0)}{\Big |}_{\rm plateau} \times \mathcal{A}_{\rm s}\,\kappa^{2(1-\alpha_{\rm s})}\times \mathcal{W}_{num}(\kappa)\,,~~~~ \kappa \equiv {k\over k_{\rm RD}} ={ f\over f_{\rm RD}}\\
 \label{eq:Wnum}
\mathcal{W}_{num}(\kappa) \equiv  \pi \kappa^2\left(a(\tau)\over a_{\rm RD}\right)^2|\mathcal{H}_k(\tau)|^2\,,\hspace*{2.5cm}
\end{eqnarray}
with $a(\tau)$ and $\mathcal{H}_k(\tau)$ the solution to the differential equations
\begin{eqnarray}\label{eq:Numcond}
a'(\tau) = a(\tau)^2 H_* \left((1 - \delta) a(\tau)^{-3 (1 + w_{\rm s})} + \delta\, a(\tau)^{-4}\right)^{1/2}\,,~~~a (\tau_*) = 1\\
\label{eq:Numcond2}
\mathcal{H}_k''(\tau) + 2{a'(\tau)\over a(\tau)}\mathcal{H}_k'(\tau) + k^2\mathcal{H}_k(\tau) = 0\,,~~\left\lbrace\begin{array}{l}
\mathcal{H}_k(\tau = \epsilon/k) = 1\,, \\\mathcal{H}_k(\tau = \epsilon/k) = 0\,,
\end{array}\right. \,,
\end{eqnarray}
where $\epsilon \ll 1$ is an arbitrary small (positive) number guaranteeing the evolution of the tensor modes to start at super-horizon scales, and $\delta \equiv {\rho_{\rm rad}^*/\rho_*}$ is the initial fraction of the radiation energy density. We observe that if we average the expression of $\mathcal{W}_{osc}(\kappa)$ from Eq.~(\ref{eq:Wosc}) over mode oscillations, we recover the expression for $\mathcal{W}(\kappa)$ from Eq.~(\ref{eq:WindowFunct}), as it should. 

\section{Detection Prospects}
\label{sec:Results}

\begin{figure}[t]
	\centering
	\includegraphics[width=7cm]{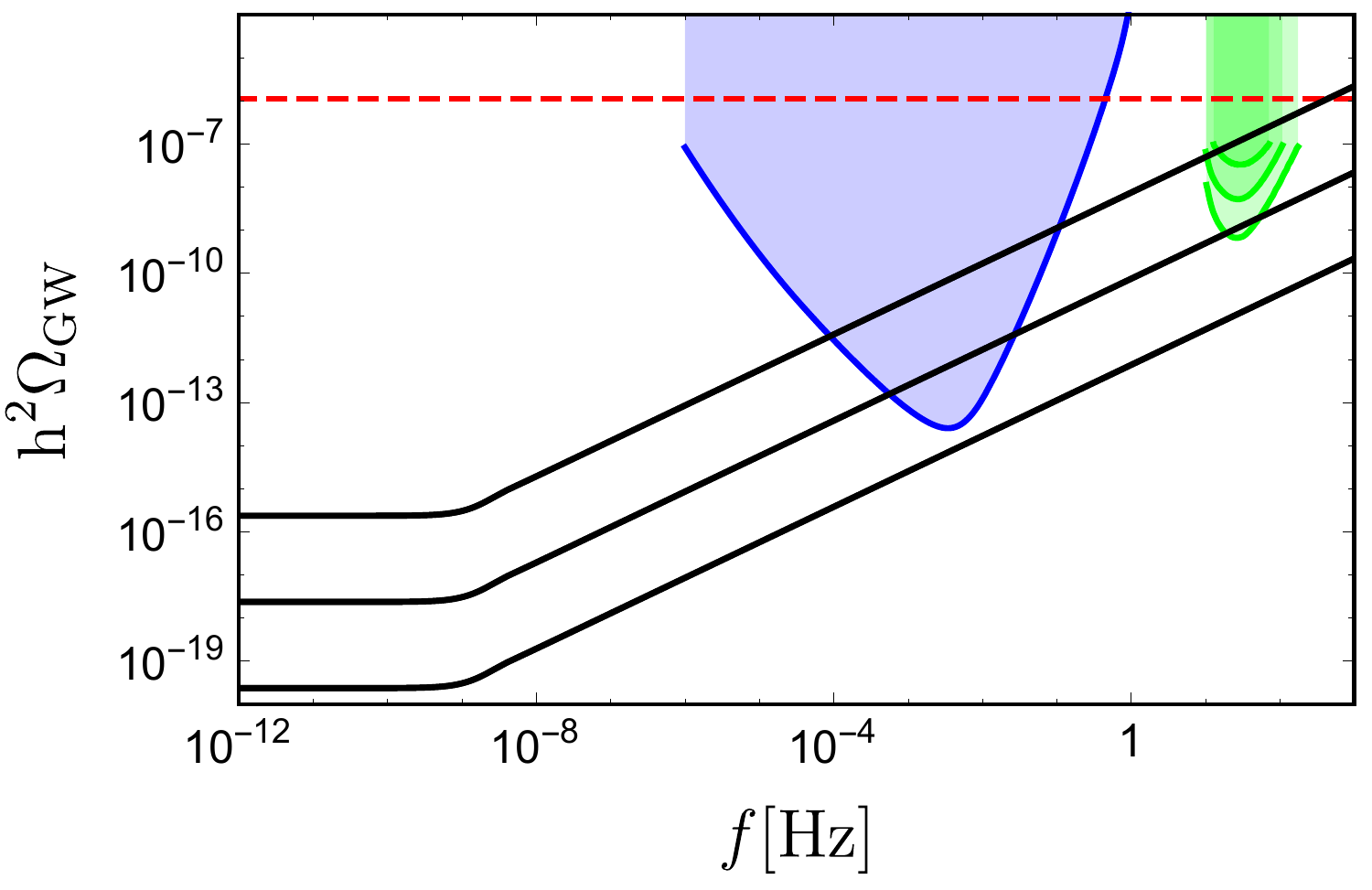}~~~~~
	\includegraphics[width=7cm]{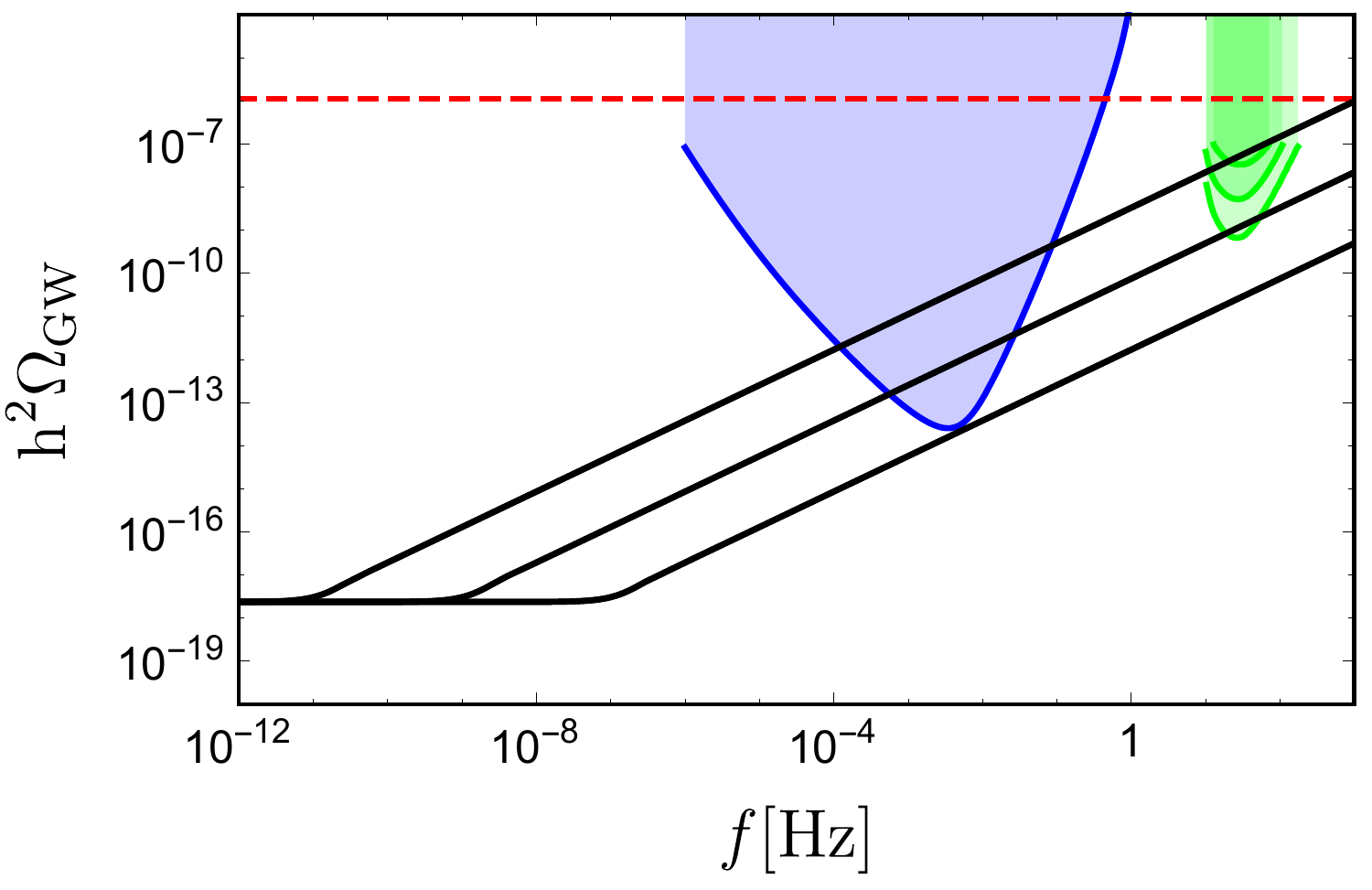}\vspace*{0.5cm}
	\includegraphics[width=7cm]{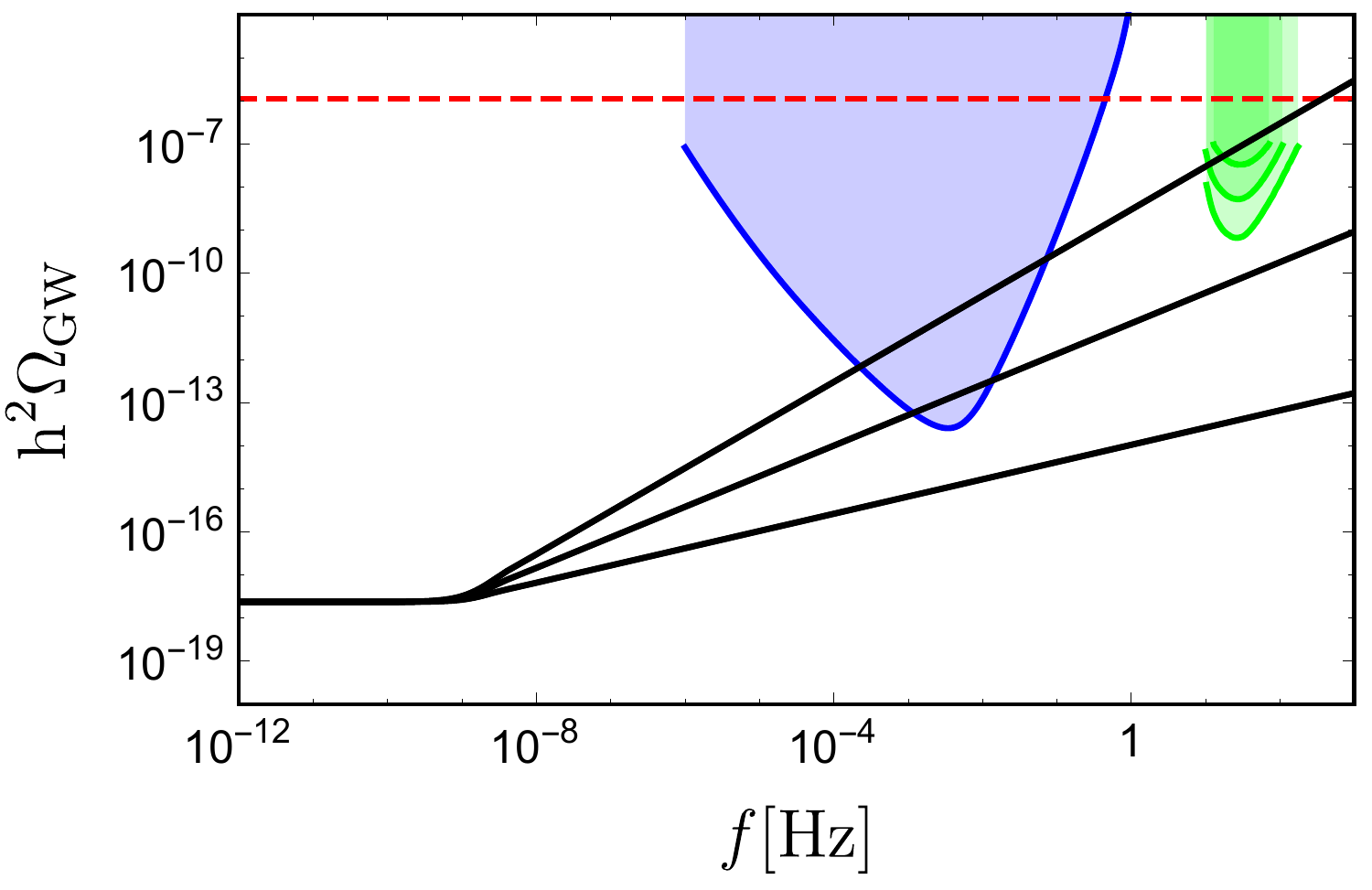}~~~~~~~
	\includegraphics[width=6.8cm,height=4.5cm]{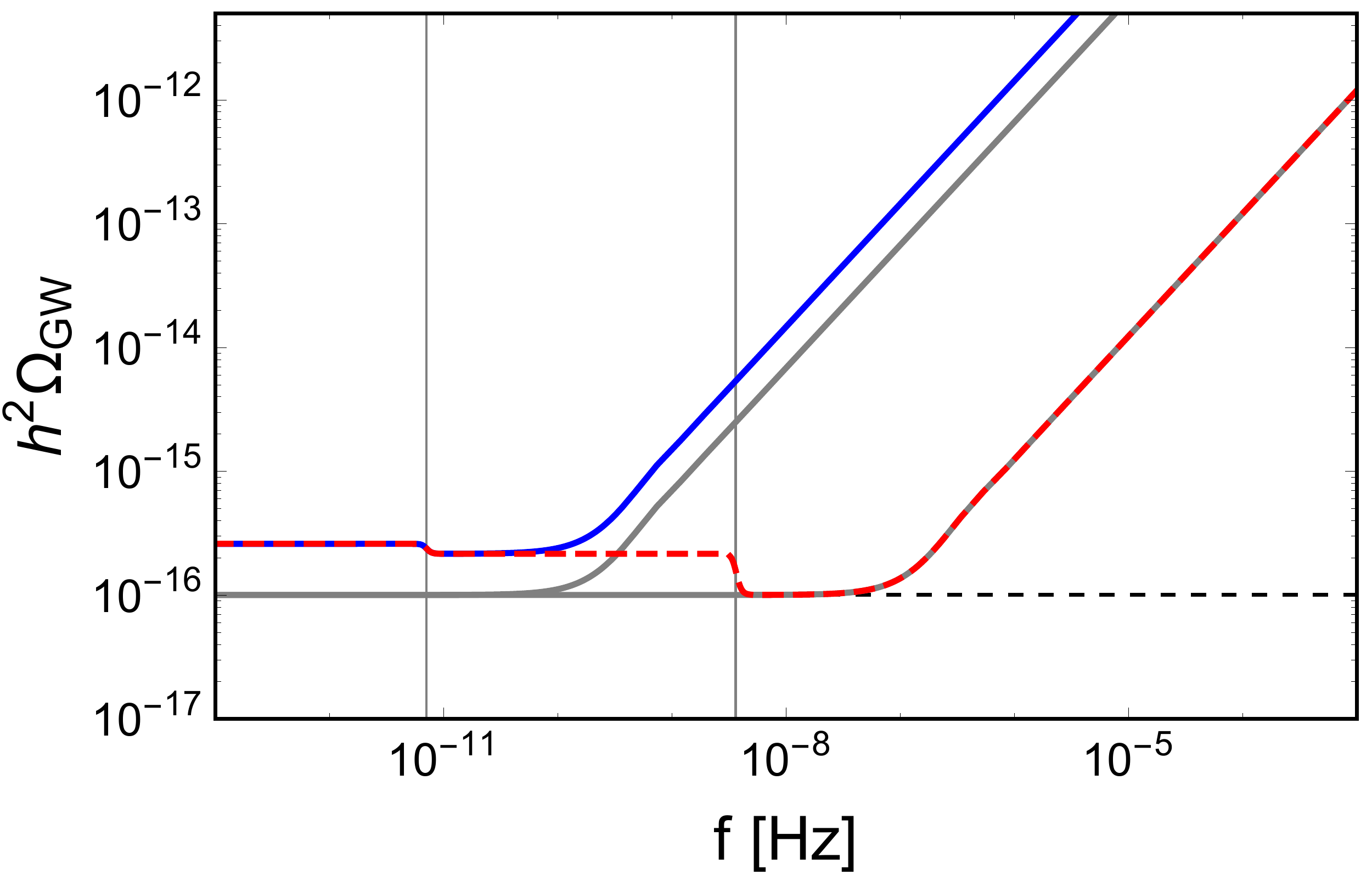}
	\caption{{\it Top-left, top-right, bottom-left panels}: GW energy spectra in the instant transition case (black solid lines), together with the LISA sensitivity curve (blue solid line), and LIGO sensitivity curves (green solid lines). We also indicate the BBN upper bound (red dotted line), which we introduce later in Sect.~\ref{subsec:BBNandCMBbounds}. In the top-left panel we fix $w_{\rm s}=0.8$ and $f_{\text{RD}}=10^{-9}\text{ Hz}$ and plot the GW energy spectra for $H_{\text{inf}}=10^{12}\text{ GeV}$, $10^{13}\text{ GeV}$, $10^{14}\text{ GeV}$. In the top-right panel we fix $H_{\text{inf}}=10^{13}\text{ GeV}$ and $w_{\rm s}=0.8$ and plot the GW energy spectra for $f_{\text{RD}}=10^{-7}\text{ Hz}$, $10^{-9}\text{ Hz}$, $10^{-11}\text{ Hz}$. In the bottom-left panel we fix $H_{\text{inf}}=10^{13}\text{ GeV}$ and $f_{\text{RD}}=10^{-9}\text{ Hz}$ and plot the GW energy spectra for $w_{\rm s}=0.5$, $0.7$, $1$. {\it Bottom-right panel}: GW energy spectra in the instant transition case taking into account changes in the number of relativistic $dof$, depending on whether the SD-to-RD transition takes place before (red dashed line) or after (blue solid line) the QCD phase transition. For comparison we show the corresponding spectra (gray solid lines) without correcting for changes in the number of $dof$. We postpone the discussion on this to Sect.~\ref{s:changesindof}.}
	\label{fig:GWenergyspectrum}
\end{figure}

Regardless of how we model the SD-to-RD transition, the overall shape of the GW energy spectrum today $h^2\Omega_{\text{GW}}(\tau_0,f)$ can be characterized by three parameters: the inflationary Hubble rate $H_{\text{inf}}$, the equation of state parameter deep inside the SD epoch $w_{\rm s}$, and the frequency today $f_{\text{RD}}$ corresponding to the horizon scale at the transition from SD to RD. The spectra exhibit in all cases a low frequency plateau at $f \ll f_{\rm RD}$, and a power law branch $\propto f^{2(1-\alpha_{\rm s})}$ at large frequencies $f \gg f_{\rm RD}$. In the top and left-bottom panels in Figure~\ref{fig:GWenergyspectrum}, we plot the oscillation averaged spectra in the instant transition case, evaluated at different values of the parameters $\lbrace w_{\rm s}, H_{\rm inf}, f_{\rm RD}\rbrace$. As can be seen in these panels, $H_{\text{inf}}$ controls the level of the plateau (and hence the overall amplitude of the full spectrum, see the left-top panel), whereas $w_{\text{s}}$ determines the slope of the high-frequency part, and $f_{\text{RD}}$ determines the location of the ``elbow" where the plateau and the blue-tilted parts are connected. In the mentioned panels in Figure~\ref{fig:GWenergyspectrum}, we also plot the power-law sensitivity curves for stochastic GW backgrounds of LISA and the O1, O2 and O5 runs of advanced LIGO. As it is evident from the figure panels, there are values of the parameters for which we expect the signal to be clearly observable by LIGO and LISA. In the following, we will first determine the parameter space $\lbrace w_{\rm s}, H_{\rm inf}, f_{\rm RD}\rbrace$ compatible with a detection and later subtract from it the region that is incompatible with current upper bounds on stochastic GW backgrounds, as set by BBN and CMB constraints.

Had we also plot in Figure~\ref{fig:GWenergyspectrum} the high-frequency branch of the GW oscillation-averaged energy spectra in the smooth-transition case, the difference between the instant- and smooth- transition spectra would be barely noticeable to the eye (given the logarithmic scales involved). For convenience, whenever we plot a GW spectrum against frequency today, we will use the instant transition case, as only then we have an analytic expression for the full spectral range, all the way from small $f \ll f_{\rm RD}$ to large $f \gg f_{\rm RD}$ frequencies, c.f.~Eq.~(\ref{eq:GWfullSpectrumInstant}). For the analysis we present in the rest of this section, however, we could use either case, as we only need to evaluate the high-frequency branch of the spectrum, for which we have the analytic expressions for both instant- and smooth- transitions, recall Eqs.~(\ref{eq:GWasympInstant}) and (\ref{OGW0}). As an instantaneous SD-to-RD transition can at best be an approximation, we consider the smooth-transition case as more realistic. For the following parameter analysis we will use therefore the smooth transition case.

To start with, we limit our parameter-space scan within the following direct bounds imposed on each of the three parameters: the non-detection of primordial B-modes in the CMB puts an upper bound on the energy scale of inflation $H_{\text{inf}} \leq H_{\rm max} \simeq 6.6\times 10^{13}\text{ GeV}$, c.f.~Eq.~(\ref{eq:Hmax}); the EoS of an SD epoch must lie within the range $1/3 < w_{\rm s} < 1$ so that it does not lead to a superluminal speed of sound; finally, the requirement to reheat the Universe before the onset of BBN sets a lower bound on the transition frequency, $f_{\rm RD} \geq f_{\rm BBN}$, where $f_{\rm BBN}$ is the red-shifted frequency today corresponding to the horizon scale at the onset of BBN, 
\begin{eqnarray}
f_{\rm BBN} &\equiv& {1\over 2\pi}{a_{\rm BBN}\over a_0}{H_{\rm BBN}\over\rm GeV}\times 1.52\cdot 10^{24}~{\rm Hz}\\ &=& {\Omega_{\rm rad}^{(0)}\over 2\pi}\sqrt{{H_0\over \rm Hz}{H_{\rm BBN}\over\rm GeV}}\times 1.23\cdot 10^{12}~{\rm Hz} \simeq 1.41\cdot 10^{-11}~{\rm Hz}\,,
\label{eq:fBBN}
\end{eqnarray}
where we have used $\Omega_{\rm rad}^{(0)} \simeq 9.15\cdot 10^{-5}$, $g_{s,{\rm BBN}} = g_{s,0} \simeq 3.91$, $g_{*,{\rm BBN}} = g_{*,0} = 3.36$, $H_0\simeq 67.8 \times 3.24\cdot10^{-20}$ Hz, and $H_{\rm BBN} = \pi\left({g_{*,{\rm BBN}}/90}\right)^{1/2}(T_{\rm BBN}^2/m_p)$, with $T_{\rm BBN} = 0.001$ GeV.

\subsection{Parameter space region probe-able by LISA and Advanced LIGO}
\label{subsec:LISAandLIGOspace}

While expressing the GW energy spectrum in terms of $f_{\text{RD}}$ yields a neat expression, it is more useful from a physical point of view to characterize the point of the SD-to-RD transition in terms of an energy scale, namely the temperature $T_{\text{RD}}$ of the radiation $dof$ at that moment\footnote{The radiation components in our case are likely to be in thermal equilibrium before the SD-to-RD transition, but this is not guaranteed. To avoid this problem, we could define instead an energy scale associated to $f_{\rm RD}$, as the $4th$ root of the radiation energy density $E_{\text{RD}} = \rho_{\rm rad}(\tau_{\rm RD})^{1/4}$ at $\tau = \tau_{\rm RD}$. However, since in the hot Big Bang picture it is customary to parametrize the energy scale of the RD era by the temperature of the thermal radiation, we will stick to this convention. We will simply characterize the energy scale by a temperature parameter $T_{\rm RD}$, related to the radiation energy density as $\rho_{\rm rad}(\tau_{\rm RD}) \equiv {\pi^2\over 30}g_{*,{\rm RD}}T_{\rm RD}^4$, regardless of whether the relativistic degrees of freedom are in thermal equilibrium. For all cases of (observational) interest, we expect that the SM degrees of freedom in the radiation bath will have reached thermal equilibrium before the transition time $\tau = \tau_{\rm RD}$.}. For a smooth SD-to-RD transition 
\begin{eqnarray}
T_{\text{RD}} &\equiv& \left({30\over g_{*,{\rm RD}}\pi^2}\rho_{\text{rad}}(\tau_{\rm RD})\right)^{1/4} = \left({30\over g_{*,{\rm RD}}\pi^2}{3\over2}m_p^2H_{\rm RD}^2\right)^{1/4}\nonumber\\
&=& \mathcal{G}_{\rm RD}^{-1/4}g_{*,{\rm RD}}^{-1/4}\times {\sqrt{2\pi}\sqrt[4]{90}\over\sqrt[4]{\Omega_{\rm rad}^{(0)}}}\sqrt{m_p\over H_0} \times 6.6\cdot10^{-25}\,\left({f_{\rm RD}\over{\rm Hz}}\right)~{\rm GeV} \nonumber\\
&\simeq& \mathcal{G}_{\rm RD}^{-1/4}g_{*,{\rm RD}}^{-1/4} \times 6.75\cdot 10^{7}\,\left({f_{\rm RD}\over{\rm Hz}}\right)~{\rm GeV}\label{ERDfRD}
\end{eqnarray}
where we have used where we have used $f_{\rm RD}=2\pi a_{\rm RD}k_{\rm RD}$, $k_{\rm RD} = a_{\rm RD}H_{\rm RD}$, \eqref{aRDa0smooth}, and in the last line $\Omega^{(0)}_{\rm rad}\simeq 9\times 10^{-5}$ and $H_0\simeq 67.8\text{ (km/s)/Mpc}$. Note that in the instant-transition case, the $rhs$ of Eq.~(\ref{ERDfRD}) would simply be multiplied by $\sqrt{2}$. Eq.~(\ref{ERDfRD}) informs us that the energy scales of, e.g.~the electroweak phase transition $\sim 100~\text{GeV}$, BBN $\sim 1\text{ MeV}$, and matter-radiation equality $\sim 0.1~\text{eV}$, correspond to redshifted frequencies today of order $\sim 10^{-7}~\text{Hz}$, $\sim 10^{-11}~\text{Hz}$, and $\sim 10^{-17}~\text{Hz}$, respectively. It also says that the typical frequencies that LISA and LIGO can probe, namely $f_{\text{LISA}}\sim 0.001\text{Hz}$ and $f_{\rm LIGO}\sim 10\text{ Hz}$, correspond to energy scales $\sim 10^5\text{\text{GeV}}$ and $\sim 10^9\text{\text{ GeV}}$, respectively. The requirement that SD ends before BBN starts limits the possible values of the energy scale to $T_{\rm RD}\gtrsim 1~\text{MeV}$.

\begin{figure}
	\centering
	~~~~~~~~LISA\hspace*{6.65cm} LIGO O2\vspace{0.3cm}
	\includegraphics[width=7.4cm]{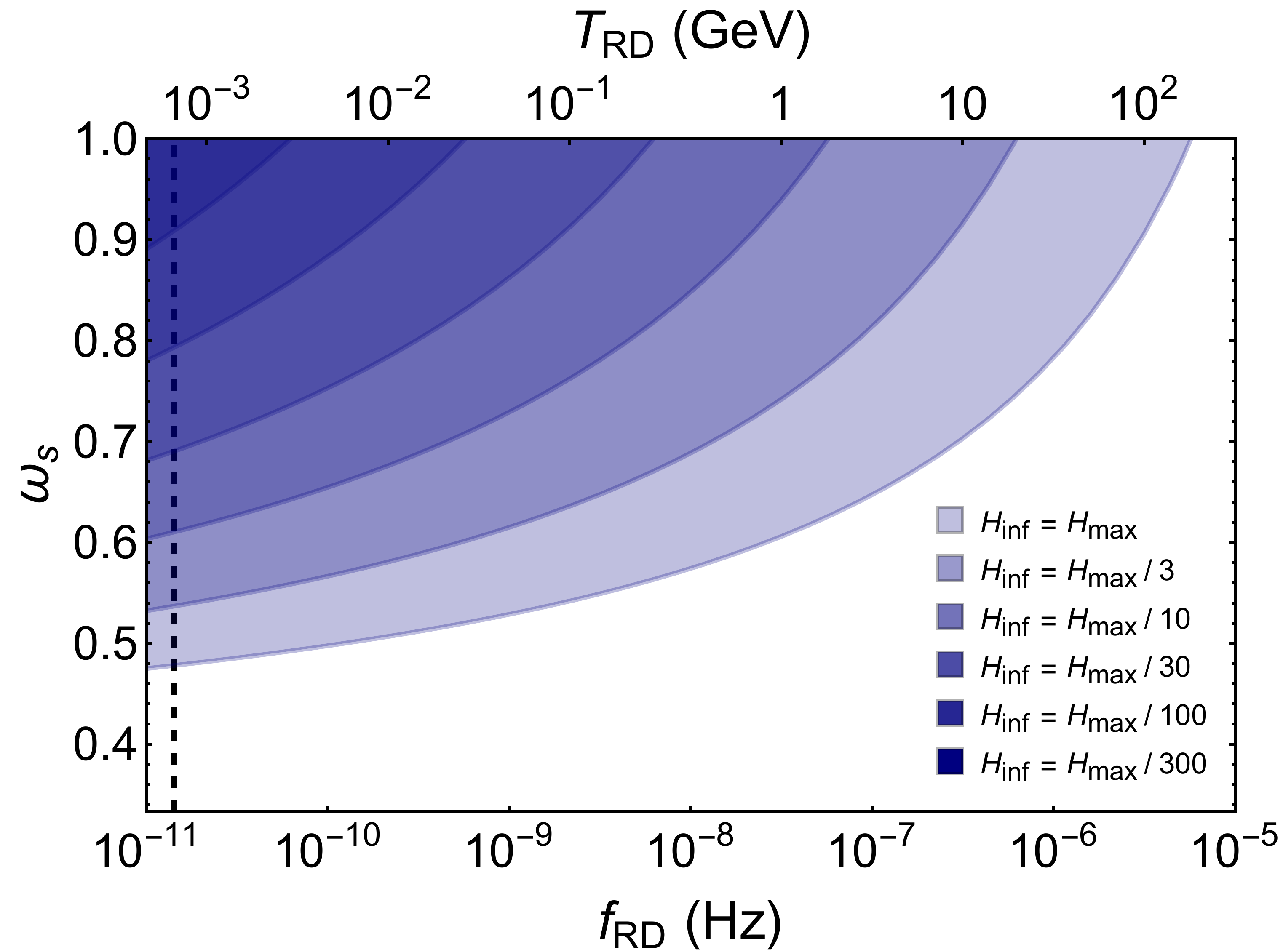}
	~~~~\includegraphics[width=7.3cm]{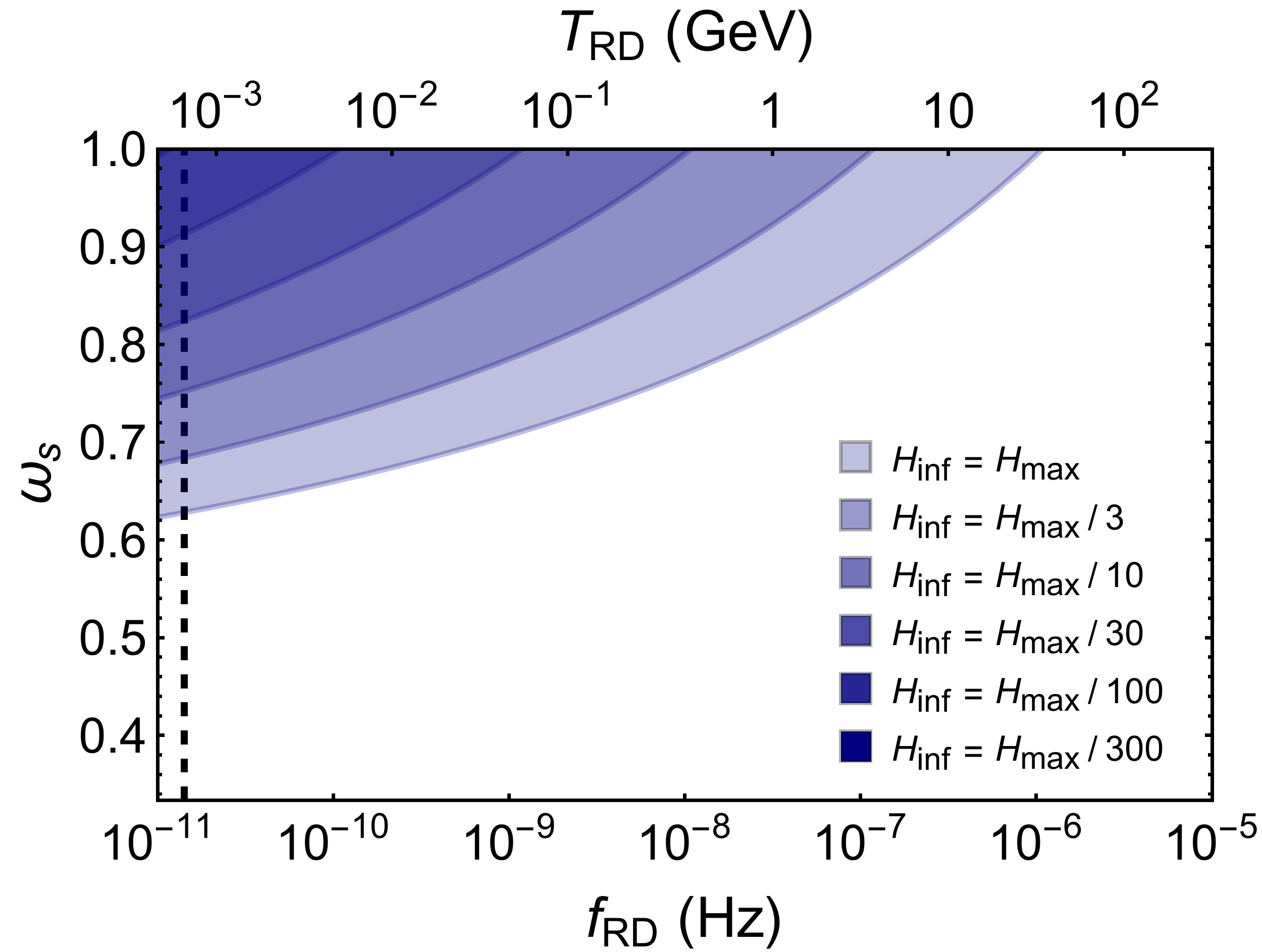}\vspace*{0.5cm}
	\hspace*{0.2cm}\includegraphics[width=7.2cm]{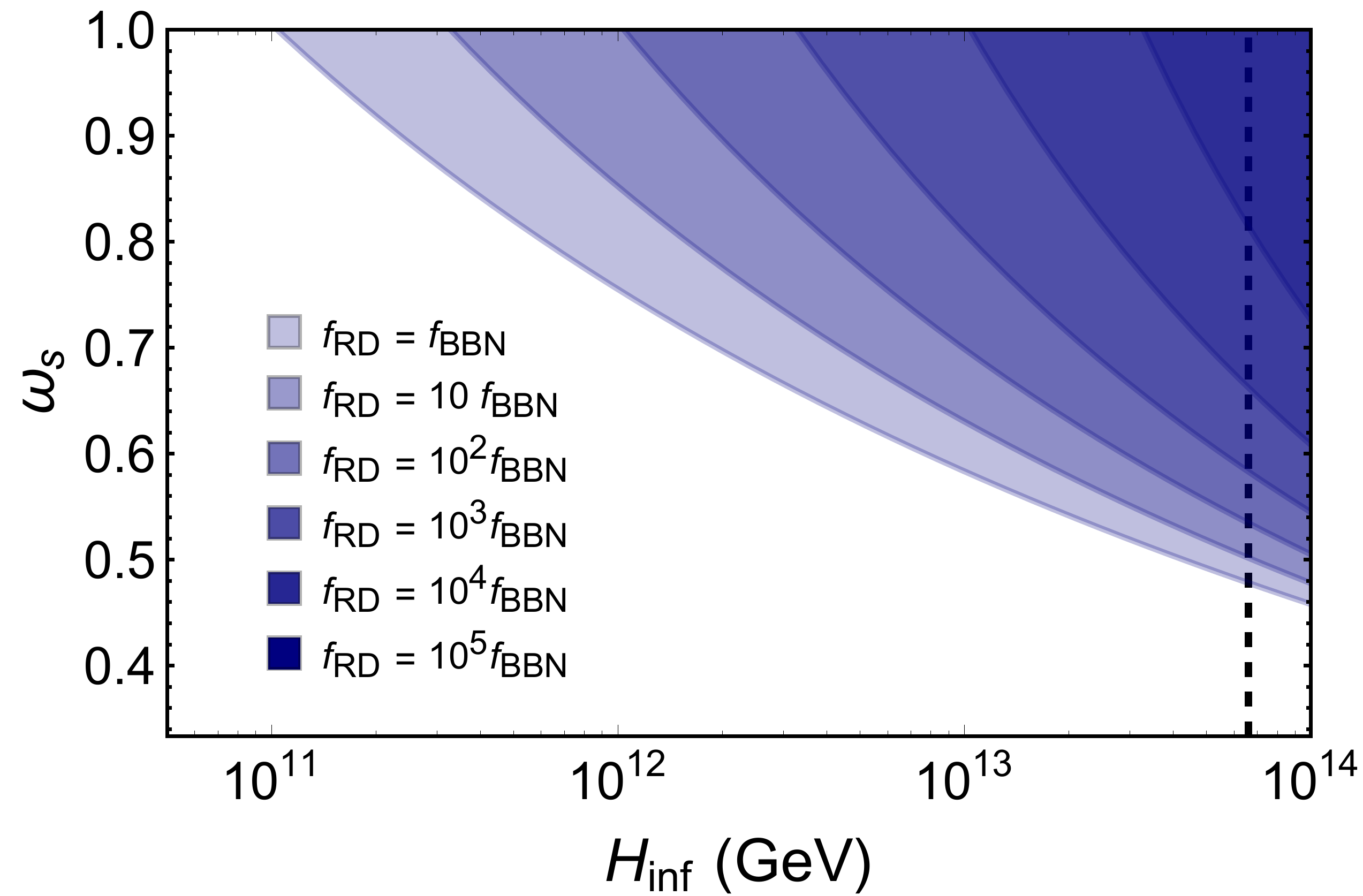}~~~~\,\,
	\includegraphics[width=7.2cm]{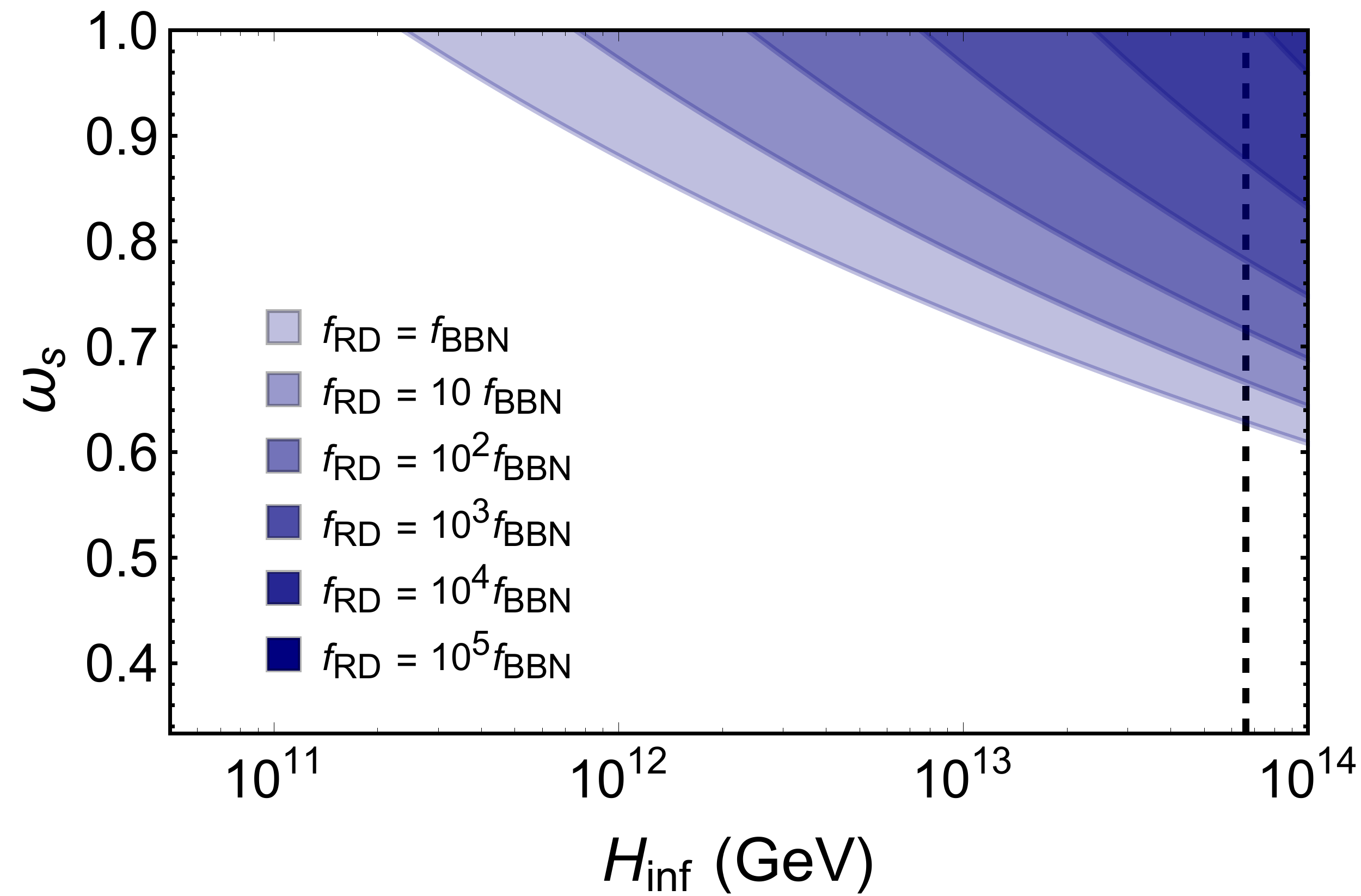}
	\caption{Colored regions represent the parameter-space probe-able by LISA (left column) and LIGO O2 run (right column). In the top row we show fixed-$H_{\rm inf}$ slices and in the bottom row fixed-$f_{\rm RD}$ slices. The vertical dashed line indicates $f_{\rm BBN}$ in the top panels, and $H_{\rm max}$ in the bottom panels. In each figure, different slices of the parameter space overlay one another, with darker ones placed upper in the stack. Beneath each slice of darker color, there are always hidden lighter colored regions.}
	\label{fig:probeableparameterspaceslices}
\end{figure}

In Figure~\ref{fig:probeableparameterspaceslices}, we present  fixed-$H_{\text{inf}}$ and fixed-$f_{\text{RD}}$ (or $T_{\text{RD}}$) slices of the parameter-space regions, probe-able by LISA and the O2 run of aLIGO. For the time being, in the present analysis, and until the end of Section~\ref{subsec:BBNandCMBbounds}, we fix $n_t = 0$ and the number of $dof$ to $g_{*,{\rm RD}} = g_{s,{\rm RD}} \simeq 106.75$. In other words, the effects of a red-tilt in the GW spectrum and of changes in the number of relativistic $dof$ in the RD epoch are not included in these plots. We will present corrections to our analysis due to the inclusion of these effects in Sections~\ref{s:spectraltilt} and \ref{s:changesindof}, respectively. In light of the left panels in Figure~\ref{fig:probeableparameterspaceslices}, we find that in order to detect the GW background by LISA, the values of the parameters must lie in the following ranges
\begin{align*}
10^{11} \text{ GeV} \lesssim ~&H_{\text{inf}} \leq 6.6\times 10^{13}\text{  GeV}\,,\\
0.48 \lesssim ~&w_{\text{s}}\phantom{_{\rm D}} < 1\,,\\
10^{-11}\text{ Hz}\lesssim ~&f_{\text{RD}} < 5.7\times 10^{-6}\text{ Hz}\,,\\
10^{-3}\text{ GeV}\lesssim ~&T_{\text{RD}} < 1.52\times 10^{2}\text{ GeV}\,,
\end{align*}
which are understood as follows. If a parameter, say $H_{\text{inf}}$, lies in the above specified range, then there are values of the other two parameters, i.e.~$w_{\rm s}$ and $f_{\text{RD}}$ ($T_{\text{RD}}$), that give rise to GW signals that are detectable by LISA. 

Let us note that as we scan the parameters $H_{\text{inf}}$, $w_{\rm s}$, and $f_{\text{RD}}$ from somewhere outside the probe-able region by LISA, the GW spectrum first intersects the LISA sensitivity curve close to the minimum point of the curve, i.e. where the sensitivity is of LISA is maximum (recall the top and left-bottom  panels in Figure~\ref{fig:GWenergyspectrum}). Since the LISA sensitivity curve is steep enough around its minimum, the first intersection point of a GW spectrum lies always very close to the tip of the LISA sensitivity curve (we have checked explicitly that this is always true by varying the parameters around their boundary values that separate detection from non-detection). Hence, based on this observation, the detectable-undetectable boundary in the parameter space is given roughly by the condition $ h^2\Omega_{\text{GW}}(f_{\text{LISA}}) \gtrsim h^2\Omega_{\text{GW}}^{\rm (LISA)}$, where $f_{\rm LISA}\simeq 3.1 \text{ mHz}$ is the frequency where LISA reaches its best sensitivity, and $h^2\Omega_{\rm GW}^{\rm (LISA)}\simeq 10^{-13}$ is the amplitude of the LISA power law sensitivity curve at such minimum. Using this, we arrive to the following condition for 'detectability' with LISA in the smooth-transition case,
\begin{equation}
\left(\frac{H_{\text{inf}}}{6.6\times 10^{13}\text{ GeV}}\right)\left(\frac{3.1 \text{ mHz}}{f_{\text{RD}}}\right)^{\frac{3w_{\text{s}}-1}{3w_{\text{s}}+1}}\gtrsim {31.6\over {\tilde A}_{\rm s}^{1/2}}\, \sim 23.6-31.6\,,\label{LISAapprox}
\end{equation}
where one should substitute $\tilde{\mathcal{ A}}_{\rm s} \longrightarrow \mathcal{A}_{\rm s}$ for the instant transition case. This rough analytic bound is useful for making quick estimates. 

\begin{figure}
\hspace*{3.3cm}LISA\hspace*{5.9cm}LIGO O2\\
\includegraphics[width=6.5cm]{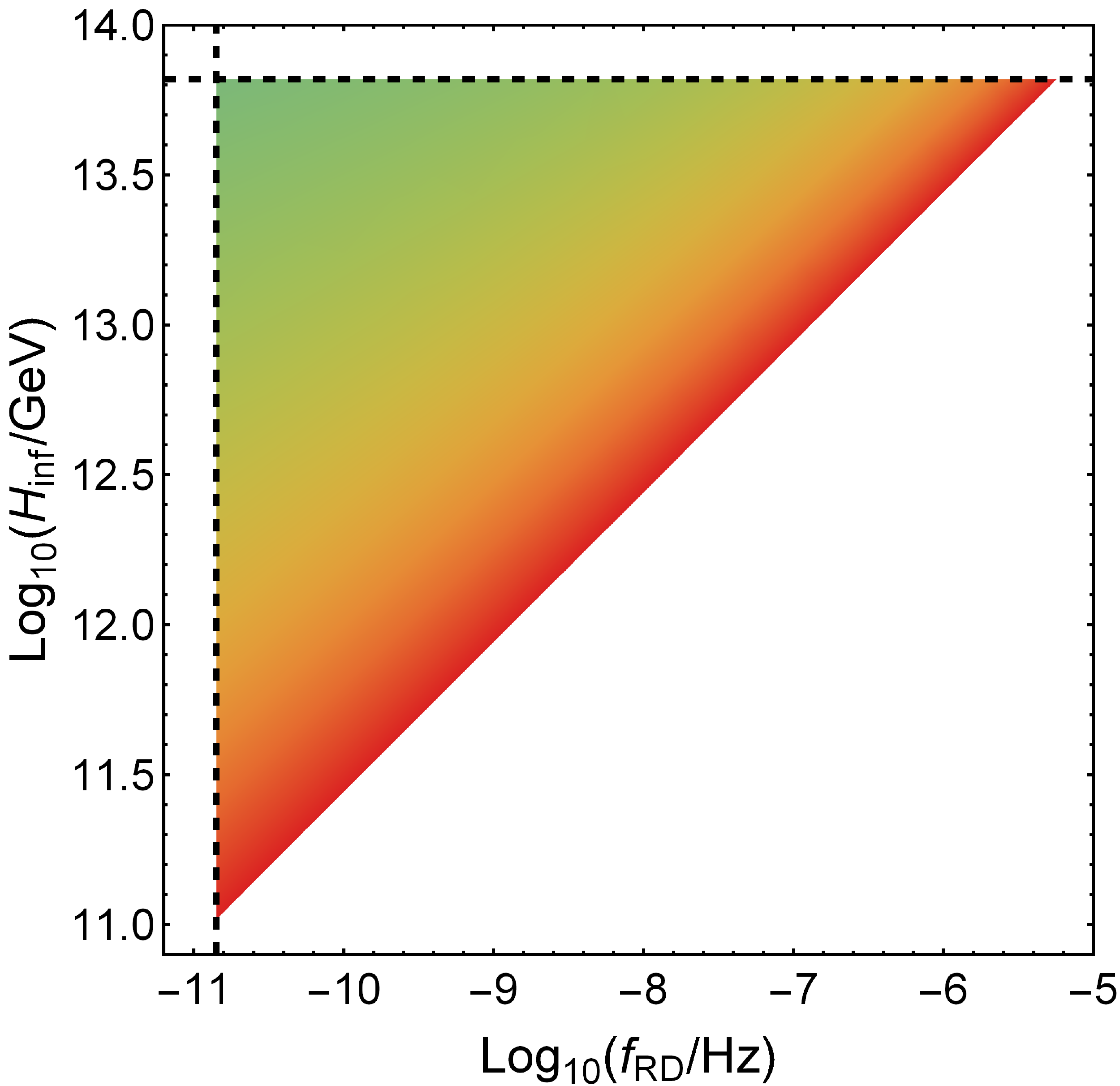}
~~~~\includegraphics[width=6.5cm]{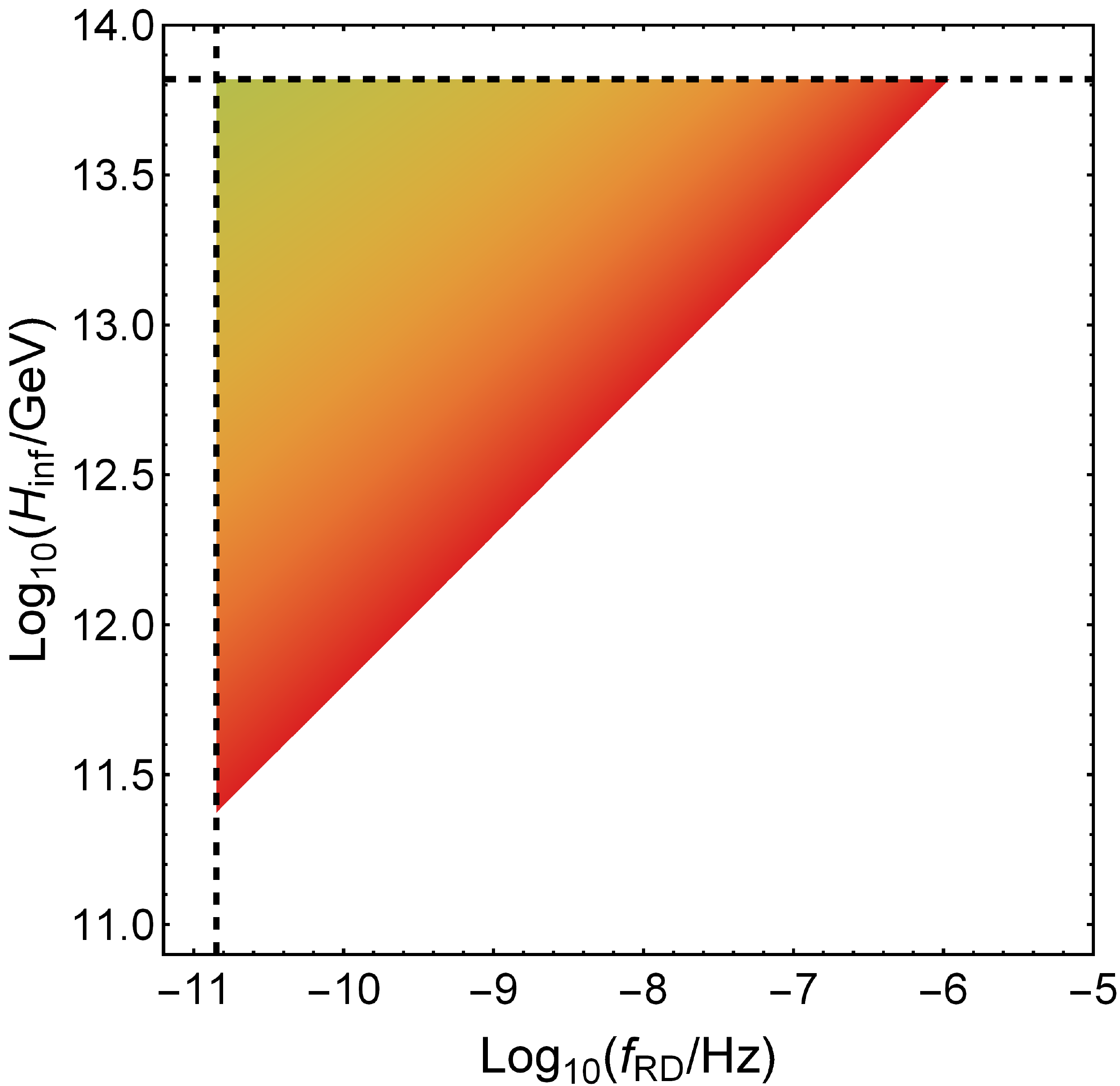}
~~~{\includegraphics[width=1cm]{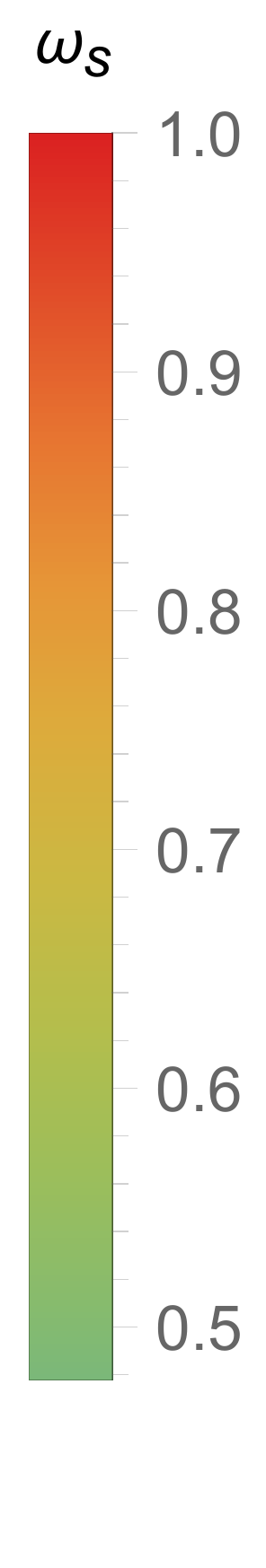}}
\caption{Parameter-space regions probe-able by LISA (left panel) and LIGO O2 (right panel), sampling gradually the value of $w_{\rm s}$. The color coding in both figures is the same, so it is clear that LISA can probe some fraction of the parameter space inaccessible to LIGO: in particular the region with sufficiently low values of $w_{\rm s}$, represented by the green-ish colors in the upper left corner of the left panel, but absent in the right panel. Vertical and horizontal dashed lines represent $f_{\rm BBN}$ and $H_{\rm max}$, respectively.}
\label{fig:probeableparameterspaceslicesII}
\end{figure}

On the other hand, as it is evident from the panels in Figure~\ref{fig:GWenergyspectrum}, whenever a GW spectrum crosses through the LISA sensitivity curve, it is possible that it will also cross through the LIGO sensitivity curve, depending on the parameters. Recently, the results from cross-correlation analysis on data from Advanced LIGO's second observing run (O2), combined with the results of the first observing run (O1), showed no evidence for the presence of a stochastic background in LIGO~\cite{LIGOScientific:2019vic} (the analysis actually focused on a search for power law GW spectra just like in our case). This implies that the parameter space accessible to aLIGO O2 should be subtracted to the parameter space probe-able by LISA that we just quantified above. The parameter space that would have led LIGO O2 to a detection of this GW background, lie in the following ranges (see right panels in Figure~\ref{fig:probeableparameterspaceslices})
\begin{align*}
2.5\times 10^{11}\text{ GeV} \lesssim ~&H_{\text{inf}} < 6.6\times 10^{13}~\text{GeV}\,,\\
0.62 \lesssim  ~&w_{\text{s}} < 1\,,\\
10^{-11}\text{ Hz}\lesssim ~&f_{\text{RD}} < 1.1\times 10^{-6}\text{ Hz}\,,\\
10^{-3}\text{ GeV}\lesssim ~&T_{\text{RD}} < 29.0\text{ GeV}\,.
\end{align*}
As can see there is still margin for a potential detection of this background by LISA, as the parameter space regions probed by LIGO O2 are in general smaller than those probed by LISA. If we were to obtain the parameter space expected to be probe-able by LIGO O5 (using the projected power law sensitivity curve), we would still reach a similar conclusion. In general, LIGO's ability to probe the GW background studied here leads to a looser minimum $H_{\rm inf}$, looser maximum $f_{\rm RD}$, and tighter minimum $w_{\rm s}$, than a detection with LISA. This is understandable since, as displayed in the top- and left-bottom panels of Figure~\ref{fig:GWenergyspectrum}, the sensitivity curves of either LIGO O2 or LIGO O5 are a factor of $\sim \mathcal{O}(10^{-5})-\mathcal{O}(10^{-4})$ less sensitive than that of LISA, but probe frequencies a factor of $\sim 10^4$ higher than LISA. Since the logarithmic slope of our the high-frequency observable branch of our GW background is constrained to be ${d\log\rho_{\rm GW}\over d\log f} = {2(w-1/3)\over (w+1/3)} < 1$, it is understandable that there is always a fraction of the parameter space that can lead to a detection in LISA, but not in LIGO (note that this never happens in the opposite direction). This is quantified in Figure~\ref{fig:probeableparameterspaceslicesII}, where we plot with the same color coding the isocurves $w_{\rm s}(H_{\rm inf},f_{\rm RD}) = const$, as we vary $H_{\rm inf}$ and $f_{\rm RD}$, both for LISA (left panel) and LIGO O2 (panel). In the figure we can clearly see that the region with sufficiently low values of $w_{\rm s}$, represented by the green-ish colors in the upper left corner of the LISA panel, is absent in the LIGO panel\footnote{Had we used the LIGO O5 expected sensitivity, the small$-w_{\rm s}$ region probe-able by LISA but inaccessible to LIGO O5 would still exist, but with a reduced volume of the parameter space.}. 

The next logical step would be to quantify the 'reduced' observable parameter space by LISA, after we subtract the parameter space region ruled out by the present non-detection of a stochastic GW background by LIGO O1+O2. Instead, we will address first in the next subsection other upper bounds on stochastic GW backgrounds, due to constraints on the expansion rate during BBN and CMB decoupling. We will find that such upper bounds are actually stronger (for the problem at hand) than the current upper bounds from LIGO O1+O2~\cite{LIGOScientific:2019vic}.

\subsection{BBN and CMB constraints}
\label{subsec:BBNandCMBbounds}

Gravitational waves contribute to the energy budget of relativistic species in the Universe. Overly abundant GWs may alter too much the expansion rate of the Universe during BBN, which, in turn, may affect the resulting abundances of light elements. To avoid sabotaging the success of BBN, it is required that \citep{Caprini:2018mtu}
\begin{equation}
\int_{f_{\text{BBN}}}^{f_{*}}h^2\Omega_{\text{GW}}(\tau_0, f)\frac{df}{f} \leq 5.6 \times 10^{-6} \, 
\Delta N_{\nu} \simeq 1.12\times 10^{-6}\,, \label{BBNbound}
\end{equation}
where $\Delta N_{\nu}$ parametrizes the extra amount of radiation beyond the SM $dof$\footnote{The contribution from extra radiation during any stage of evolution of the Universe is typically parametrized in terms of an effective deviation $\Delta N_{\nu}$ from the number of SM neutrino species $N_{\nu}=3.04$. This is only a parametrization. The extra radiation does not need to be neutrinos and can be either bosonic or fermionic.}~\cite{Mangano:2005cc}, where we have used the CMB constraint on the number of extra relativistic species $\Delta N_\nu \lesssim 0.2$ at $95\%~C.L.$~\cite{Cyburt:2015mya}. In Eq.~(\ref{BBNbound}), the lower bound $f_{\rm BBN}$ is the frequency today corresponds to the mode crossing the horizon at the onset of BBN, c.f.~Eq.~(\ref{eq:fBBN}), whereas the upper bound $f_{*}$ is the frequency today corresponds to the mode re-entering the horizon at the end of inflation (hence it is the high-frequency end of the GW spectrum). The BBN constraint above is an integral constraint, i.e.~a non-local constraint in the frequency space. To get a local, slightly weaker constraint out of it, we can simply perform the integration in \eqref{BBNbound}. As our signal is simply a power-law at high frequencies $\Omega_{\rm GW} \propto f^{2(1-\alpha)}$, the integral is dominated by its upper bound, and we obtain
\begin{equation}
h^2\Omega_{\text{GW}}(f_{*}) \lesssim (1-\alpha_{\rm s})\cdot2.24\times 10^{-6}\,,\label{localBBNbound}
\end{equation}
where the $lhs$ should be evaluated using Eq.~(\ref{eq:GWasympInstant}) for the instant-transition case, or with Eq.~(\ref{OGW0}) for the smooth-transition case. Thus, the BBN constraint tends to rule out cases with a high inflationary scale $H_{\text{inf}}$, large EoS $w_{\text{s}}$, and low transition frequencies $f_{\text{RD}}$ (equivalently low energy scales $T_{\text{RD}}$); these parameters would yield an amplitude of $h^2\Omega_{\text{GW}}(f_{*})$ that is well above the $rhs$ of Eq.~(\ref{localBBNbound}). The BBN bound  may therefore rule-out the highest possible GW signals that the detectors may probe at their respective frequency range.

In order to obtain the BBN bound on the parameter space $\lbrace H_{\text{inf}}, w_{\rm s}, f_{\text{RD}}\rbrace$, we need to compute the ratio of $f_{*}$ to $f_{\rm RD}$. This will tell us about the maximum value of $h^2\Omega_{\text{GW}}(f)$ and whether it exceeds the BBN bound. By construction, we have
\begin{align}
\frac{f_{*}}{f_{\rm RD}}=\frac{k_{*}}{k_{\rm RD}}=\frac{a_{*}H_{*}}{a_{\rm RD}H_{\rm RD}}\,.
\end{align}
To proceed, we express $a_{*}/a_{\rm RD}$ in terms of $H_{*}$, $H_{\rm RD}$, and $\alpha_{\rm s}$ using \eqref{astiff} and \eqref{aratioHratio} in the smooth transition case. Then, we express $H_{\rm RD}$ in terms of $f_{\rm RD}$, $\Omega_{\rm rad}^{(0)}$, and $H_0$ using $f_{\rm RD}=k_{\rm RD}/(2\pi a_0)$, $k_{\rm RD}=a_{\rm RD}H_{\rm RD}$, and \eqref{aRDa0smooth} in the smooth-transition case (alternatively \eqref{aRDa0abrupt} in the instant-transition case). The result is
\begin{equation}
\frac{f_{*}}{f_{\rm RD}}=\begin{cases}
\displaystyle\left[\frac{\left(\mathcal{G}_{\rm RD}\Omega_{\text{rad}}^{(0)}\right)^{1/2}H_0H_{*}}{4\pi^2 f_{\text{RD}}^2}\right]^{\frac{1}{1+\alpha_{\rm s}}}, &\text{ abrupt transition}\\
\displaystyle 2^{\frac{1}{2}\left(\frac{1-\alpha_{\rm s}}{1+\alpha_{\rm s}}\right)}\left[\frac{\left(\mathcal{G}_{\rm RD}\Omega_{\text{rad}}^{(0)}\right)^{1/2}H_0H_{*}}{4\pi^2 f_{\text{RD}}^2}\right]^{\frac{1}{1+\alpha_{\text{s}}}}, &\text{ smooth transition}
\end{cases} \label{finffRD}
\end{equation}
The relative strength of the high-frequency end of the GW spectra in the two cases is thus
\begin{equation}
\frac{\Omega_{\rm GW}^{(\rm smooth)}(\tau_0,f_{*}^{(\rm smooth)})}{\Omega_{\rm GW}^{(\rm abrupt)}(\tau_0,f_{*}^{(\rm abrupt)})}=2^{2\left(\frac{1-\alpha_{\rm s}}{1+\alpha_{\rm s}}\right)}\,,
\end{equation}
where the ratio ranges from $1$ for $w_{\rm s} = 1/3$ to $2^{2/3} \simeq 1.59$ for $w_{\rm s} = 1$.

\begin{figure}[t]
	\centering
	\includegraphics[width=7.4cm]{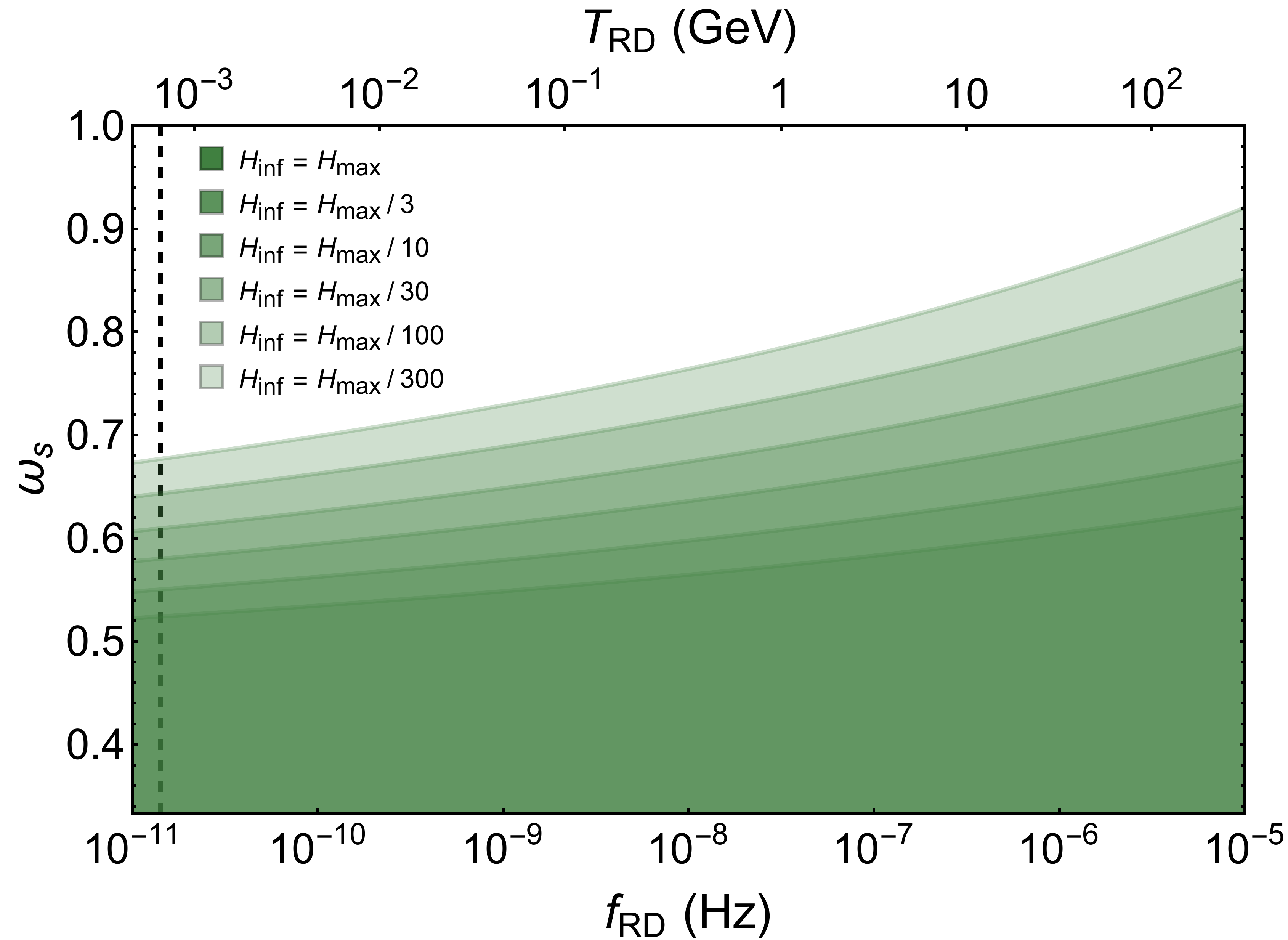}
	~~~~\includegraphics[width=7.4cm]{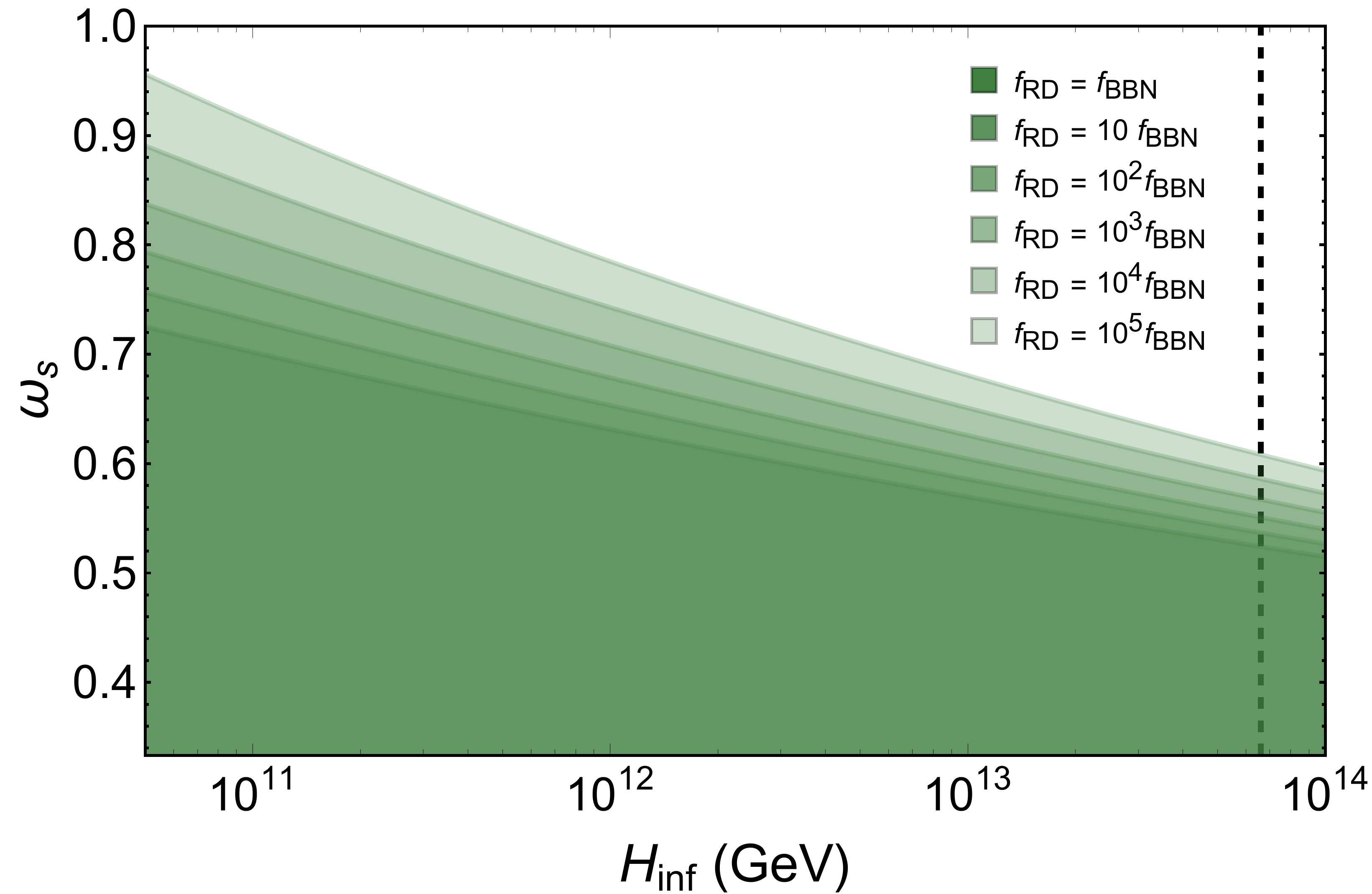}
	\caption{Colored regions represent the parameter space compatible with the requirement of not having too much GWs at the onset of BBN (to obtain these contours we have used the smooth transition GW spectrum Eq.~(\ref{OGW0}) evaluated at $f_{*}$, assuming  $g_{*,{\rm} RD} = g_{s,{\rm} RD} = 106.75$ at the SD-to-RD transition). In the left panel we show fixed-$H_{\rm inf}$ slices of the parameter space, whereas in the right panel fixed-$f_{\rm RD}$ slices. In each figure, different slices of the parameter space overlay one another, with darker ones placed lower in the stack. Beneath each slice of darker color, there are always hidden lighter colored regions.}
	\label{fig:ruledoutbyBBN}
\end{figure}

\begin{figure}[t]
~~Probe-able region compatible with BBN ~~~ $|$ ~~~ Probe-able region compatible with CMB
\centering
\includegraphics[width=7.4cm]{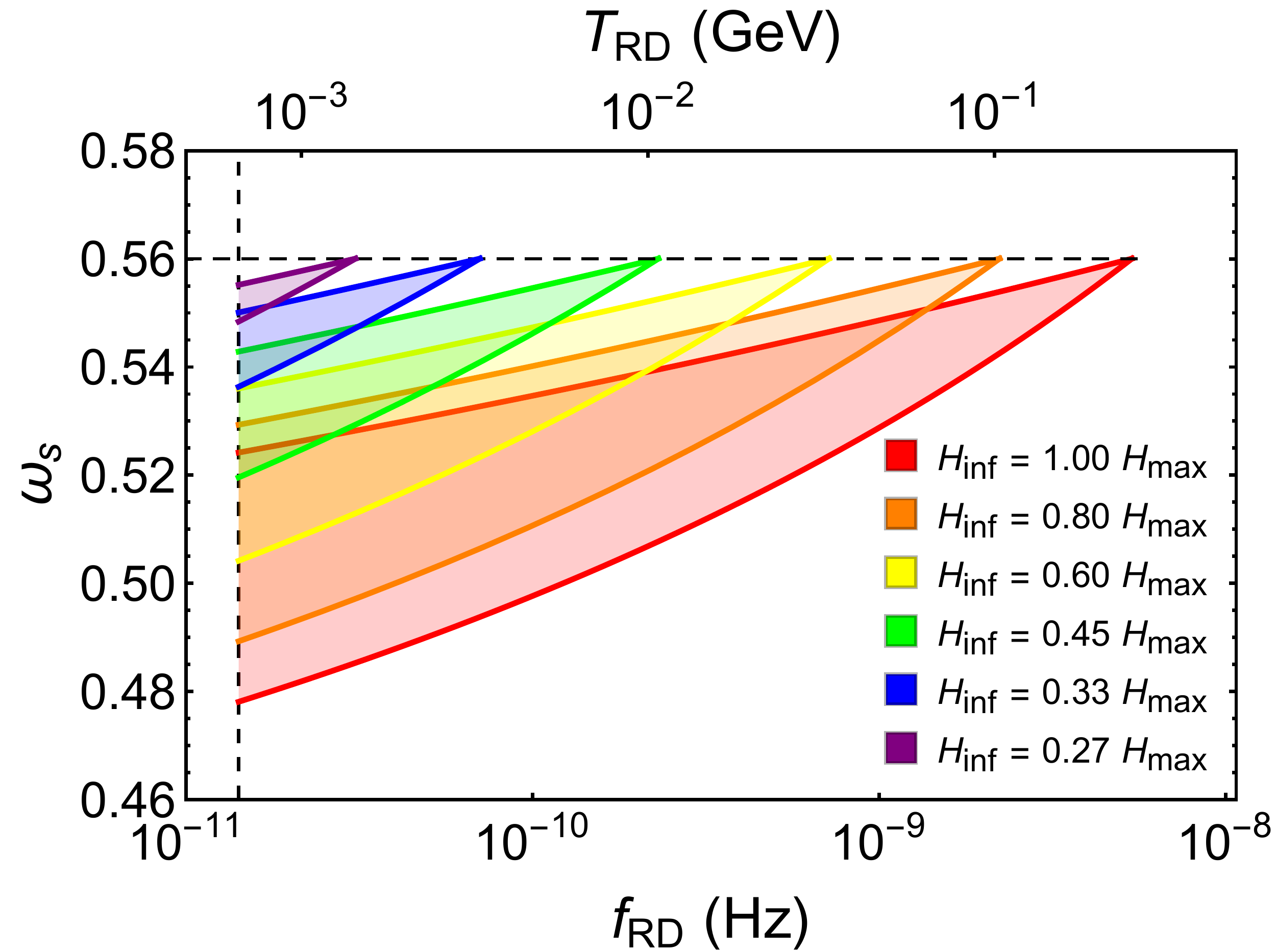}~~~
\includegraphics[width=7.4cm]{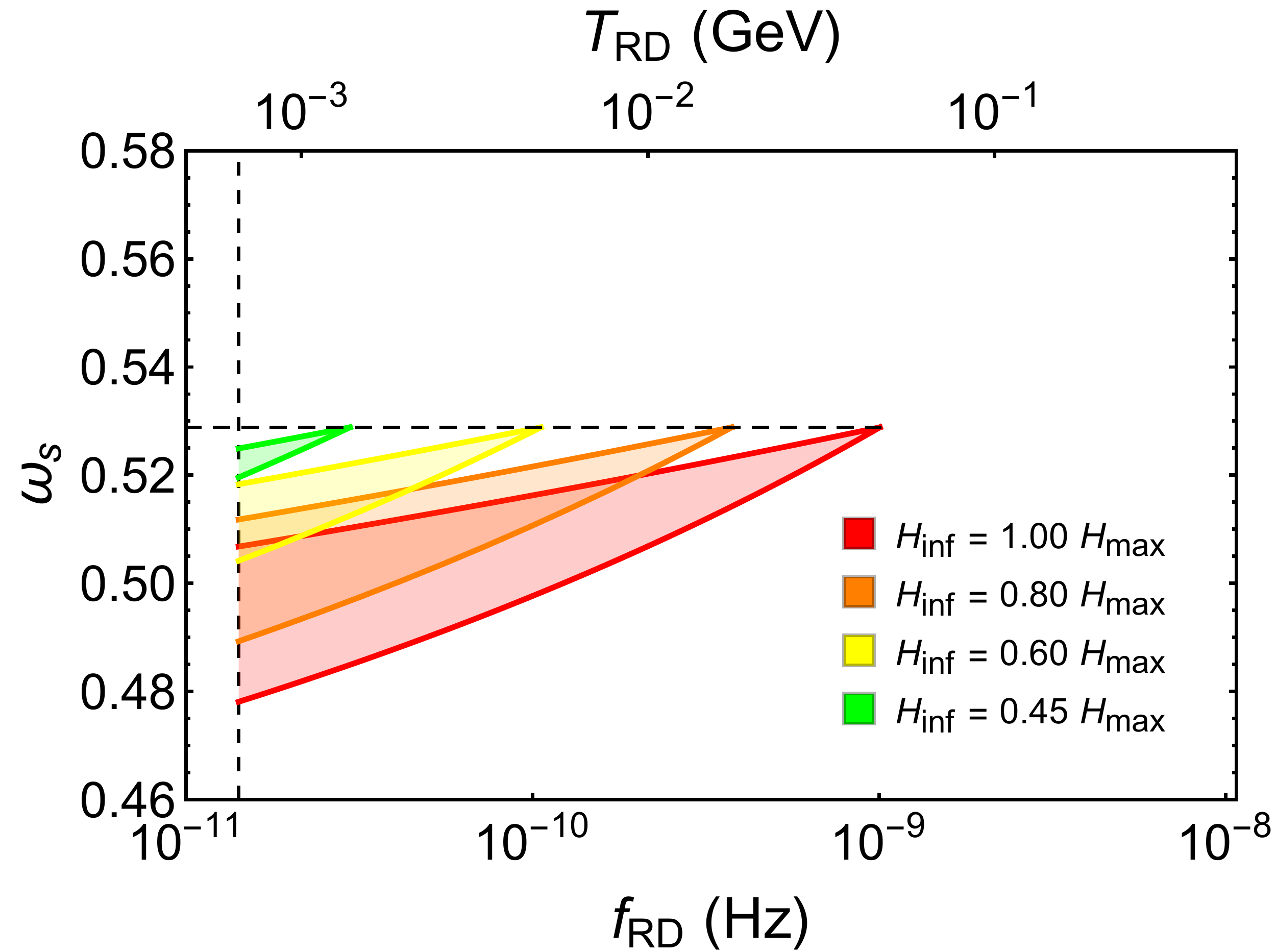}
\caption{Remaining parameter space region probe-able by LISA after removing the part that is incompatible with the upper bounds on GW stochastic backgrounds. In the left panel we show fixed-$H_{\rm inf}$ slices of the parameter space compatible with the BBN constraint Eq.~(\ref{BBNbound}), whereas in the right panel we show the analogous plot but imposing the more restrictive CMB constraint Eq.~(\ref{CMBupperbound}). If we vary $H_{\rm inf}$ continuously, instead of discretely as done here, the spiky feature of the remaining parameter space region would become infinitely dense, and the resulting figures on the left and right panels would have clean, horizontal cuts at $w_{\rm s}\approx 0.56$ and $w_{\rm s}\approx 0.53$ respectively, both which are shown above in dashed lines.}
\label{fig:remainingLISAregion}
\end{figure}

For simplicity, from now on we will make the assumption that the Hubble parameter during inflation $H_{\rm inf}$ is approximately equal to that at the end of inflation $H_*$, i.e. $H_{\rm inf}\simeq H_*$. Using \eqref{finffRD}, we can obtain the BBN bound \eqref{localBBNbound} in terms of $H_{\text{inf}}$, $w_{\rm s}$, and $f_{\text{RD}}$, and work out the parameter space region compatible with the BBN bound. The latter is shown in Figure~\ref{fig:ruledoutbyBBN} for the smooth transition case. If we exclude the region  incompatible with the BBN bound from the parameter space region probe-able by LISA shown in Figure~\ref{fig:probeableparameterspaceslices}, only a relatively small region remains, see left panel of Figure~\ref{fig:remainingLISAregion}. On the other hand, the parameter space region probe-able by current LIGO O2, or even by the will-be LIGO O5 expected sensitivity is completely wiped out when we consider the BBN bound. In other words, there is no possible value of $H_{\text{inf}}$, $w_{\rm s}$, or $f_{\text{RD}}$, that can lead to a detection in LIGO without violating the BBN bound. This implies that the inflationary background of GWs we are studying here is expected to be undetectable by current or future runs of LIGO. To put it differently, if a stochastic signal with power law spectrum was to be discovered by LIGO in the coming years, it should not be identified with the inflationary GW background in the presence of an SD era.

Interestingly, we see in the left panel of Figure~\ref{fig:remainingLISAregion} that the BBN bound appears to rule out the region $w>0.56$, essentially independently of $f_{\rm RD}$ and $H_{\rm inf}$. To understand this, let us recall that the parameter space region probe-able by LISA can be described to a good approximation by \eqref{LISAapprox}. In terms of this approximate bound, the cut shown in the Figure~\ref{fig:remainingLISAregion} is the intersection between the LISA bound saturation surface $h^2\Omega_{\rm GW}(\tau_0,f_{\rm LISA}; H_{\rm inf}, f_{\rm RD}, w_{\rm s}) = h^2\Omega_{\rm GW}^{\rm (LISA)}$, and the BBN bound saturation surface, $h^2\Omega_{\rm GW}(\tau_0,f_{*}; H_{\rm inf}, f_{\rm RD}, w_{\rm s}) = h^2\Omega_{\rm GW}^{\rm (BBN)}$. Incidentally, both $h^2\Omega_{\rm GW}(\tau_0,f_{\rm LISA})$ and $h^2\Omega_{\rm GW}(\tau_0,f_{*})$ depend on exactly the same combination of $H_{\rm inf}$ and $f_{\rm RD}$, namely $H_{\rm inf}^2/f_{\rm RD}^{2(1-\alpha_{\rm s})}$. Note that this is only true because, as stated earlier, we assumed $H_{\rm inf}\simeq H_*$. Consequently, we can combine the two equations in such a way that the dependence on $H_{\rm inf}^2/f_{\rm RD}^{2(1-\alpha_{\rm s})}$ drops, leaving behind an equation for $\alpha_{\rm s}$ (or $\omega_{\rm s}$). This explains why the BBN bound cuts in the left panel in Figure~\ref{fig:remainingLISAregion} appears as a straight line, independent of $H_{\rm inf}$ and $f_{\rm RD}$. 

A similar bound on the amount of GWs can also be obtained from constraints on the Hubble rate at CMB decoupling, as this can be used to infer an upper bound on extra radiation components~\cite{Smith:2006nk,Sendra:2012wh,Pagano:2015hma}. This translates to an upper bound on the amount of GWs, which actually extends to a greater frequency range than the BBN bound, down to $f \lesssim 10^{-15}$ Hz. In Ref.~\cite{Smith:2006nk} two cases were identified, depending on the initial conditions of the GWs. In the first case, labeled as the ‘adiabatic initial condition’, the GW background is assumed to have perturbations imprinted on its energy density following the same distribution as all other components in the Universe. In the second case, labeled as the ‘homogeneous initial condition’, the GW background is not perturbed and the curvature perturbation is the one of the standard adiabatic case. We view this second option
for initial conditions as more justified, since it applies to most of the known cosmological GW backgrounds, and certainly to the background studied in this paper. The most recent analysis on this~\cite{Pagano:2015hma} only analyzes the case of adiabatic initial conditions. Extrapolating this result to the case of GWs with homogeneous initial conditions, Ref.~\cite{Caprini:2018mtu} concludes that the constraint should yield approximately a bound as
\begin{equation}
h^2\Omega_{\rm GW}(\tau_0, f)\lesssim 2\times 10^{-7}\,, \label{CMBupperbound}
\end{equation}
which is a factor $\sim 5$ more stringent than the BBN bound~\eqref{BBNbound}. In the right panel of  Figure~\ref{fig:remainingLISAregion}, we show how the parameter space region probe-able by LISA would change if we take into consideration the CMB upper bound~\eqref{CMBupperbound} on the amount of GWs. The CMB bound \eqref{CMBupperbound} suggests that there must be less GWs in the Universe (in the form of a stochastic background that permeates all space) than suggested by the BBN bound \eqref{localBBNbound}. Consequently, the parameter space compatible with a detection by LISA and satisfying at the same time Eq.~(\ref{CMBupperbound}) is smaller than the parameter space compatible with a detection by LISA while satisfying Eq.~(\ref{BBNbound}) [compare the size of the areas of the parameter regions in the right panel of Figure~\ref{fig:remainingLISAregion} with respect to those in the left panel]. Identical reasoning as before explains as well the straight line cut in the right panel in Figure~\ref{fig:remainingLISAregion} which rules out the region $w_{\rm s} \gtrsim 0.53$, independently of $H_{\rm inf}$ and $f_{\rm RD}$. The CMB bound is however not as robust as the BBN bound, because  Eq.~(\ref{CMBupperbound}) is based on a extrapolation of the actual CMB constraint, based on an adiabatic initial condition for the GW background. We therefore prefer to stick only the BBN bound \eqref{localBBNbound} in what follows.

From the left panel of Figure~\ref{fig:remainingLISAregion} it becomes clear that the initially large regions of parameter space compatible with a signal detection by LISA, recall left panels in Figure~\ref{fig:probeableparameterspaceslices}, has shrunk into a small region where the values of the parameters lie in the following ranges
\begin{align*}
6.6\times 10^{12}\text{ GeV} \lesssim ~&H_{\text{inf}} < 6.6\times 10^{13}~\text{GeV}\,,\\
0.48 \lesssim ~&w_{\text{s}}\phantom{_{a}} < 0.56\,,\\
10^{-11}\text{ Hz}\lesssim ~&f_{\text{RD}} \lesssim 5.5\cdot 10^{-9}\text{ Hz}\,,\\
1\text{ MeV}\lesssim ~&T_{\text{RD}} \lesssim 215\text{ MeV}\,.
\end{align*}

\subsection{Further effects that may modify the GW energy spectrum}
\label{subsec:TiltAndDof}
{subsec:BBNandCMBbounds}
There are other physical effects capable of distorting the shape of the GW energy spectrum that we have not taken into account in the analysis so far. These effects have not been included in the main analysis because they induce only small deviations in the results, and depend on some unknown aspects of the physics at high energies. In this section we study the impact of adding a small red-tilt in the GW spectrum, the spectral distortion due to changes in the number of relativistic $dof$ during RD, and an anisotropic stress due to the presence of free-streaming relativistic neutrinos. 

\subsubsection{Red spectral tilt }\label{s:spectraltilt}

Due to the fact that the inflationary phase cannot be perfectly {\it de~Sitter}, we should consider the effect of a red-tilt in the spectrum of GWs. This is expected e.g.~in basic slow-roll inflationary scenarios. We can easily incorporate such tilt into our analysis, by simply multiplying the previous GW energy spectra with a factor
\begin{equation}
\Omega_{\rm GW}(\tau_0, f)\rightarrow \Omega_{\rm GW}(\tau_0, f)\times \left(\frac{f}{f_{\rm p}}\right)^{n_{\rm t}}\,,
\end{equation}
where $f_{\rm p}$ is a pivot frequency (related to CMB scales). The spectral tilt is constrained as $-n_{\rm t}\lesssim 0.008$~\cite{Akrami:2018odb,Ade:2018gkx}, so it changes the tensor spectrum only very gradually with scale. As mentioned before, in the absence of running of the spectral index, i.e.~if ${dn_t\over d\log k} = 0$, the amplitude of the tensor spectrum only falls by a factor $\sim (10^{26})^{-0.008} \sim 0.6$ during the $\ln(e^{60}) \sim $ 26 orders of magnitude separating the CMB scales from the Hubble radius at the end of inflation. This is what justifies our assumption that the low-frequency branch $f \ll f_{\rm RD}$ of the spectrum is scale-invariant. To quantify the effect in our analysis due to adding a tilt in the spectrum, we choose the same pivot scale used in the Planck series of papers, given by a present day wavenumber 
\begin{equation}
\frac{k_{\rm p}}{a_0}\simeq 0.002\text{ Mpc}^{-1}\,,
\end{equation}
which corresponds to a frequency today as 
\begin{equation}
f_p=\frac{1}{2\pi}\frac{k_{\rm p}}{a_{0}}\simeq 3.09\times 10^{-18}\text{ Hz}\,.
\end{equation}
The left panel of Figure~\ref{figure:tensortiltcorrected} shows the remaining parameter space region that is probe-able by LISA when imposing a red spectral tilt $n_{\rm t}=-0.008$ in the GW spectrum. Due to the weakening of the GW spectrum, we expect that slightly lower values of $H_{\rm inf}$, higher values of $f_{\rm RD}$, and smaller values of $w_{\rm s}$ to be no longer probe-able by LISA. At the same time, we also expect the BBN bound to loosen, allowing higher values of $w_{\rm s}$ to be probe-able. Such shifting effects of the parameter space regions are clearly visible in the left panel of Figure~\ref{figure:tensortiltcorrected}, where the probe-able regions in the presence of tilt (colored areas, delimited by solid lines) are displaced upwards (i.e.~to higher values of $w_{\rm s}$) and towards the left (i.e.~to smaller values of $f_{\rm RD}$), when compared to the probe-able regions in the absence of tilt (empty areas, delimited by dashed lines). Overall, the probe-able regions still represent very small islands of the parameter space. As expected, the presence of a tilt (even in the maximum case as we chose) results in only a small correction.

\subsubsection{Changes in the effective number of relativistic $dof$ during RD}\label{s:changesindof}

While the energy density of the GWs always scales as $a^{-4}$, the total energy of the universe during the RD period at temperature $T$
\begin{equation}
\rho_{\text{tot}}=\frac{\pi^2}{30}g_*T^4\,,  
\end{equation}
may evolve in more complicated ways due to possible changes in the effective number $g_*$ of relativistic degrees of freedom ($dof$). In general, the temperature evolves according to entropy conservation
\begin{equation}
s=\frac{2\pi^2}{45}g_{s}T^3\propto a^{-3} ~~ \Rightarrow ~~ T \propto g_s^{-1/3}a^{-1}\,,
\end{equation}
with $g_s$ the number of entropic $dof$. Thus, the radiation energy density evolves as
\begin{equation}
\rho_{\text{tot}}\propto g_*g_{s}^{-4/3}a^{-4} \sim g_*^{-1/3}a^{-4}\,,
\end{equation}
where in the last equality we have taken into account the fact that energy and entropy $dof$s are approximately equal. If there is a sudden change in the number of $dof$, say, from $g_{*1}$ to $g_{*2}$, $\rho_{\text{tot}}$ would change by a factor of $\sim (g_{*1}/g_{*2})^{1/3}$, and consequently $\Omega_{\text{GW}} \equiv  (d\rho_{\text{GW}}/d\ln f)/\rho_{\text{tot}}$ will change by a factor of $\sim (g_{*2}/g_{*1})^{1/3}$. Note that the change in $\Omega_{\text{GW}}$ only affects the modes that are already sub-horizon before the change $g_{*1}\longrightarrow g_{*2}$ takes place; those that are super-horizon remain frozen. After a series of changes in the relativistic $dof$, the net change in $\Omega_{\rm GW}$ for a particular mode $k$ depends only on the value at the moment of horizon re-entry $g_{*,k}\equiv g_{*}(k = aH)$. As a rule of thumb, modes that re-enter the horizon earlier get a larger suppression, since the temperature and hence $g_{*,k}$ were larger then. The dependence on the number of relativistic $dof$ is encoded in the pre-factor $\mathcal{G}_k$ [c.f.~Eq.~\ref{eq:TransferAndGk}] in the amplitude of the GW spectrum [recall Eq.~(\ref{eq:InfGWtodayRD})], which we rewrite here for convenience, 
\begin{equation}\label{deltagstarRD}
\Omega_{\text{GW}}(\tau_0,k) \propto \mathcal{G}_k\,,~~~ 
\mathcal{G}_k \equiv \left(g_{*,k}\over g_{*,0}\right)\left(g_{s,0}\over g_{s,k}\right)^{4\over3}\,.
\end{equation}

\begin{figure}[t]
	\centering
	\includegraphics[width=7.4cm]{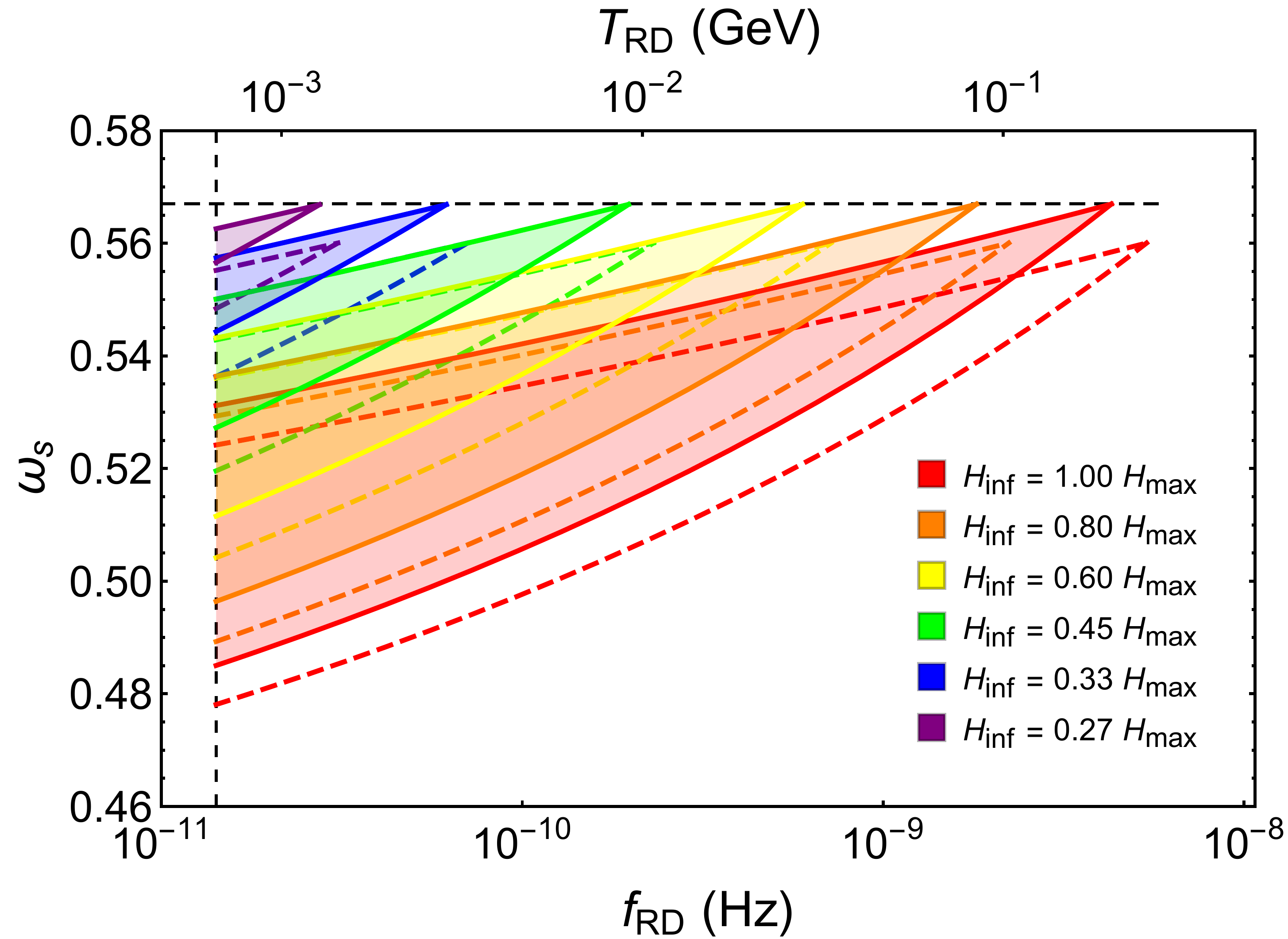}~~
	\includegraphics[width=7.4cm]{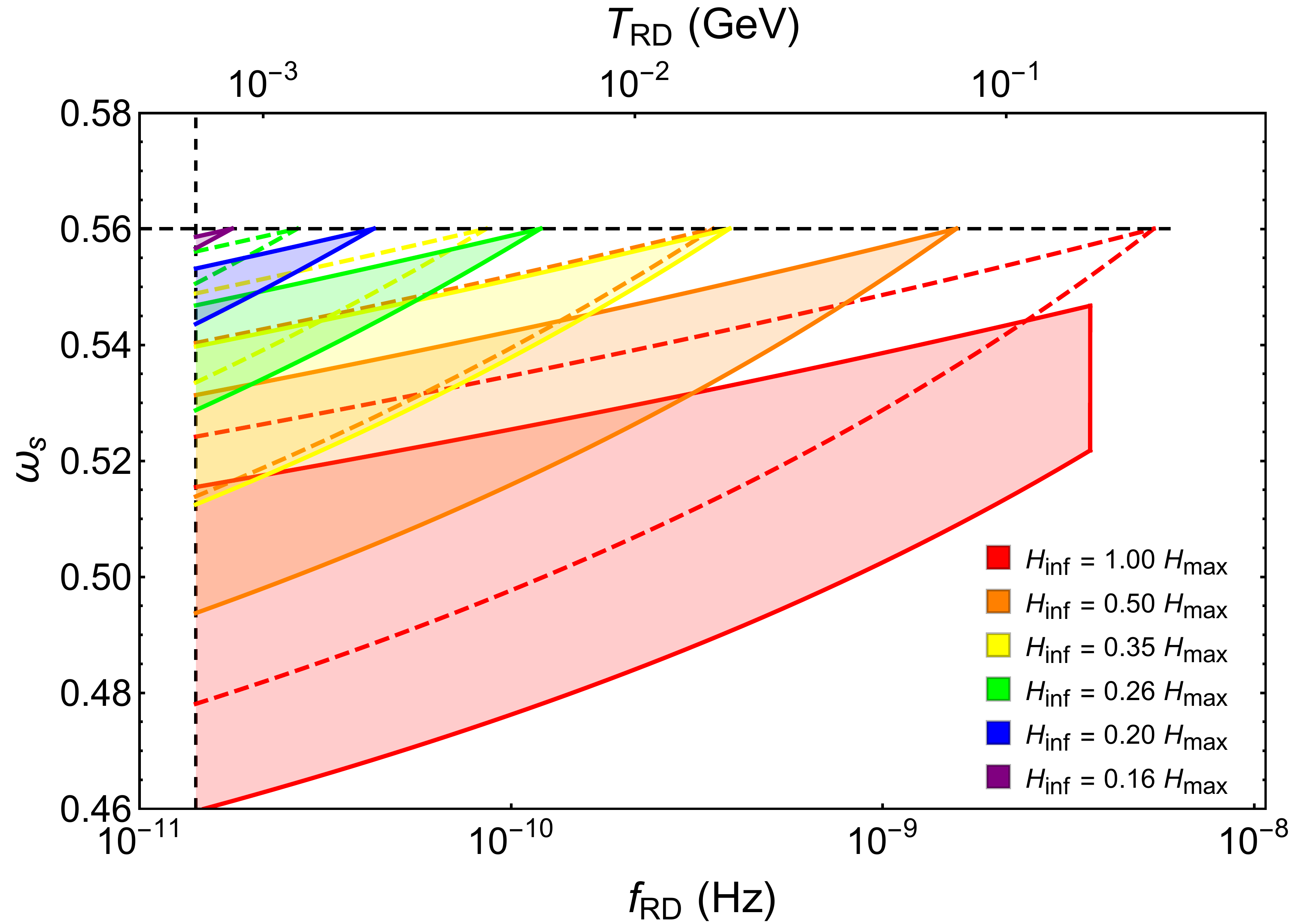}
	\caption{The remaining parameter space probe-able by LISA, accounting for the presence of a red spectral tilt $n_t = -0.008$ (left panel), and accounting for the effect of changes in the number of relativistic $dof$ (right panel). In both panels, we show for comparison as empty regions delimited by dashed line contours the probe-able regions before we considered the presence of a spectral tilt or corrected by the pre-factor $C_{\Delta g_*}$.}
	\label{figure:tensortiltcorrected}
\end{figure}

As far as the Standard Model is concerned\footnote{Changes in beyond the SM $dof$ can be also probed by LISA~\cite{Caldwell:2018giq}, but here we focus on the SM $dof$ only.}, there are only two events that lead to significant changes (in terms of percentage) in the relativistic $dof$ that we need to pay attention to. They are the QCD phase transition at around the temperature $T_{\text{QCD}}\sim 200\text{ MeV}$, and the electron-positron annihilation at a temperature $T_{e^+e^-}=0.5\text{ MeV}$. So far we simply fixed $g_{*,k} = g_{s,k} = 106.75$, which is valid whenever all the SM particles are relativistic and dominate the energy budget of the Universe. We evaluated the $plateau$ amplitude Eq.~(\ref{eq:InfGWtodayRD}) using  such values for the number of $dof$, which lead to a suppression $\mathcal{G}_k \simeq 0.39$. After the QCD phase transition, however, the number of $dof$ drop to $g_{*,k} = g_{s,k} = 10.75$, so that the suppression is smaller, with $\mathcal{G}_k \sim 0.83$. After electron-positron annihilation and neutrino decoupling, the number of $dof$ drop to their final values $g_{*,k} = g_{*,0} = 3.36$ and $g_{s,k} = g_{s,0} \simeq 3.91$, so that there is no suppression any more as $\mathcal{G}_k = 1$ (the neutrinos are treated as if they were massless because they are relativistic during the radiation era, and give a negligible contribution to the total energy density during the matter era). Therefore, to account for the effect of changes in the relativistic $dof$ in the RD epoch, we simply need to modify the main result as follows 
\begin{equation}
\Omega_{\text{GW}}(\tau_0,k)\rightarrow C_{\Delta g_*}(k)\Omega_{\text{GW}}(\tau_0,k)\,,~~~ C_{\Delta g_*}(k) \equiv {\mathcal{G}_k(g_{*,k},g_{s,k})\over \mathcal{G}_k(106.75,106.75)}\,. 
\end{equation}
A simple approximation for the correction $C_{\Delta g_*}(k)$ is to consider it as a piece-wise constant function, discontinuous at every moment the number of $dof$ changes significantly,  i.e.
\begin{equation}\label{eq:cases3}
C_{\Delta g_*}(k)=\begin{cases}
1.00\,, & k > k_{\rm QCD}\\
2.15\,, & k_{\rm QCD} > k > k_{e^+e^-}\\
2.59\,, & k < k_{e^+e^-}\\
\end{cases}\,,
\end{equation}
with $k_{\rm QCD}, k_{e^+e^-}$ the comoving horizon scales at the QCD phase transition and electron-position annihilation, respectively. Here we have implicitly assumed that the QCD phase transition happened during the RD epoch, i.e.~$T_{\text{RD}} > T_{\text{QCD}}$ (equivalently $k_{\rm RD} > k_{\rm QCD}$). In the bottom-left panel of Figure~\ref{fig:GWenergyspectrum}, we show an example of a GW spectrum (red dashed line) taking into account the correction (\ref{eq:cases3}). For comparison we show in the same panel (gray solid line) the GW spectrum for the same parameters but without the correction $C_{\Delta g_*}$.

If the change in $g_*$ happens during the SD epoch, the correction needs however to be modified. During the SD period, a jump in the value of $g_*$ in the (sub-dominant) radiation sector does not entail a jump in $\Omega_{\text{GW}}$.
Therefore, if the QCD phase transition happened during the SD epoch\footnote{We always assume that electron-positron annihilation occurs in the RD epoch, as the minimum reheating temperature required for a consistent cosmology is $T_{\rm RH} \gtrsim T_{\rm BBN} \sim$ MeV.}, i.e.~if $T_{\text{RD}} < T_{\text{QCD}}$, the correction factor is then changed to
\begin{equation}\label{eq:cases2}
C_{\Delta g_*}(k)=\begin{cases}
2.15 &\text{for } k >  k_{e^+e^-}\\
2.59 &\text{for } k <  k_{e^+e^-}\\
\end{cases}\,,
\end{equation}
which reflects that in this case, only changes in the number of $dof$ taking place during the RD, impact on the final form of $\Omega_{\text{GW}}(\tau_0,k)$. In the bottom-left panel of Figure~\ref{fig:GWenergyspectrum}, we show an example of a GW spectrum (blue solid line) taking into account the correction (\ref{eq:cases2}), and for comparison we also show the GW spectrum for the same parameters but without the correction (gray solid line). The pre-factors $C_{\Delta g_*} \gtrsim 1$ in (\ref{eq:cases2}) indicate that in the case of a late SD-to-RD transition at low energy scales, slightly before BBN, we had suppressed too much the GW spectrum by assuming that $g_{\rm RD} \sim 106.75$. In these cases, the overall amplitude of the GW energy spectra for the frequency range of interest ($f \gg f_{e^+e^-} \sim 10^{-11}$ Hz) is in reality a factor $C_{\Delta g_*} \sim 2$ larger than what we had assumed (see blue solid line, compared to the gray line, in the bottom-left panel of Figure~\ref{fig:GWenergyspectrum}). In the right panel of Figure~\ref{figure:tensortiltcorrected}, we show the comparison between the remaining region probe-able by LISA with or without the $C_{\Delta g_*}(k)$ factor taken into account. As the probe-able temperature scales correspond all to small values $T \lesssim T_{\rm QCD}$, it is important that we apply the prescription given by Eq.~(\ref{eq:cases2}), and not by (\ref{eq:cases3}). This represents therefore a relatively important correction to the GW spectra with such low $f_{\rm RD}$ frequencies, as we were extra-suppressing their amplitude (in the frequency range of interest) by a factor $\sim 2.15$. The impact of this in our parameter constraint analysis, is that the remaining probe-able regions by LISA are displaced to the right (towards larger values of $f_{\rm RD}$) and are slightly wider in $w_{\rm s}$, see colored regions in the right panel of Figure~\ref{figure:tensortiltcorrected}, which now extend the probe-able region down to $w_{\rm s} \simeq 0.46$. The probe-able parameter space ranges change now to
\begin{align*}
10^{11} \text{ GeV} \lesssim ~&H_{\text{inf}} \leq 6.6\times 10^{13}\text{  GeV}\,,\\
0.46 \lesssim ~&w_{\text{s}}\phantom{_{\rm D}} < 0.56\,,\\
10^{-11}\text{ Hz}\lesssim ~&f_{\text{RD}} < 3.6\times 10^{-9}\text{ Hz}\,,\\
1\text{ MeV}\lesssim ~&T_{\text{RD}} < 150 \text{ MeV}\,.
\end{align*}
Overall, this effect is still a small correction, which does not change the fact that after subtraction of the parameter space region incompatible with the BBN bound, the remaining parameter space probe-able by LISA is still a very small island of the full parameter space. Finally we note that correcting our previous GW spectra by a factor $\sim 2.15$ still does not change the fact that there is no surviving parameter space compatible with having a signal detection by LIGO O2 or O5, without violating the BBN constraint.

\subsubsection{Free-streaming of relativistic particles}

Free-streaming decoupled relativistic particles moving along their geodesics back-react to the spacetime in an anisotropic way, which manifests itself as an anisotropic contribution to the stress-energy tensor \citep{Weinberg:2003ur,Watanabe:2006qe}. An extra term appears then in the right hand side of the equation of motion of tensor perturbations~\eqref{eom}. If the free-streaming species in question are stable at time scales much longer than the instantaneous Hubble time, their energy density $\rho_{\text{FS}}$ make up an approximately constant fraction $f_{\text{FS}}$ of the total energy density of the universe during RD. In that case, their anisotropic contribution to the stress energy tensor can be calculated explicitly \citep{Watanabe:2006qe}
\begin{equation}
\Pi_k=-4\rho_{\text{FS}}(\tau)\int_{\tau_{\nu\text{ dec}}}^{\tau}d\tau'\left\{\frac{j_2\left[k(\tau-\tau')\right]}{k^2(\tau-\tau')^2}\right\}h_k^\prime(\tau^\prime)\,,
\end{equation}
so that \eqref{eom} becomes an integro-differential equation
\begin{equation}
h_k^{\prime\prime}+2\frac{a^\prime}{a}h_k^\prime+k^2h_k=-24\Omega_{\text{FS}}\left(\frac{a'}{a}\right)^2\int_{\tau_{\nu\text{ dec}}}^{\tau}d\tau'\left\{\frac{j_2\left[k(\tau-\tau')\right]}{k^2(\tau-\tau')^2}\right\}h_k^\prime(\tau^\prime)\,.
\end{equation}
The fact that the kernel on the right hand side can be written as a sum of spherical Bessel functions suggests that there are solutions that can be expressed as a sum of spherical Bessel functions. Keeping the first five non-vanishing terms of such sum provides a solution with an error of less than $0.1\%$, see~\citep{Boyle:2005se} for details. In practice,  the correction translates into multiplying the sub-horizon limit of the GW spectrum by a damping factor $|A|^2 < 1$, relative to that in the absence of free-streaming particles. The damping factor $|A|^2$ ranges from 1 to 0.35, corresponding to $\Omega_{\text{FS}}=0$ and $\Omega_{\text{FS}}=1$, respectively. 

This damping effect is however local in frequency space. In the case of the SM neutrinos, for example, the damping only applies to modes with the present frequency range $f_{\rm eq} < f < f_{\nu\text{ dec}}$, i.e.~those that crossed the horizon after neutrino decoupling at $T \lesssim 1$ MeV, but before matter-radiation equality. Therefore, at the frequency window that LISA and Advanced LIGO can probe (and for this matter any other direct detection GW interferometric experiment), the GW spectrum under study is not affected by the damping effect due to free-streaming of SM neutrinos. Hence, this is not an effect that we need to incorporate into our analysis. 

\section{Summary and Outlook}

\label{sec:Discussion}

In this paper we have studied how a period characterized by a stiff equation of state $1/3<w_{\text{s}}<1$, spanning from the end of inflation to the onset of RD, impacts on the GW background from inflation. Due to this SD period, the GW energy density spectrum acquires a blue tilt $n_t \equiv {d\log \rho_{\text{GW}}\over d\log f} = 2{(w_{\rm s}-1/3)\over (w_{\rm s}+1/3)}$ in the frequency regime $f \gg f_{\rm RD}$ associated to the modes that re-enter the horizon during the stiff epoch (here $f_{\rm RD}$ is the frequency today corresponding to the horizon scale at the time of the SD-to-RD transition). As a result, the energy spectrum in the high frequency domain $f \gg f_{\rm RD}$ is considerably amplified relative to the (quasi-)scale invariant part in the low frequency domain $f \ll f_{\rm RD}$, corresponding to the modes re-entering the horizon during RD.

To obtain an exact expression for the spectrum around the scales reentering during the SD-to-RD transition, we have done a matching in the frequency space between the (high frequency) modes that reenter during SD, and the (low frequency) modes that reenter during RD. This requires the use of the exact time-dependence of the scale factor, Hubble rate and equation of state of the Universe, and a careful treatment of the evolution of the modes crossing the horizon along the transition itself. We have obtained the exact transfer function of the GW energy density spectrum around the frequencies $f \sim f_{\rm RD}$, both numerically and (whenever possible) analytically, considering both 'instant' and smooth modelings of the SD-to-RD transition, see Eqs.~(\ref{eq:WindowFunct}) and (\ref{eq:Wnum}). We find that the high frequency branch of the spectrum in the smooth case is a factor $2^{1-\alpha_{\rm s}}$ larger than in the instant case, where $\alpha_s = 2/(1+3w_{\rm s})$, recall Eq.~(\ref{OGW0}). We consider a smooth transition as more realistic, since in this case all background quantities (the scale factor, the Hubble rate and the equation of state) change smoothly and continuously along the SD-to-RD transition. Because of this we decided to perform our parameter analysis only using the GW spectra obtained for a smooth transition case. Using the GW spectrum from the instant-transition case, which assumes a sudden jump in the equation of state, leads in any case to only marginal changes in the analysis results.

The shape of the GW spectrum is controlled by $w_{\rm s}$, $f_{\rm RD}$, and the energy scale of inflation $H_{\rm inf}$. We have determined the parameter space compatible with a detection of this signal by Advanced LIGO and LISA. See Figures~\ref{fig:probeableparameterspaceslices} and \ref{fig:probeableparameterspaceslicesII}, which exhibit that, $a~priori$, a large region of the parameter space is potentially observable. Consistency with upper bounds on stochastic GW backgrounds, due to constraints on the expansion rate during BBN and CMB decoupling, rules out however a significant fraction of the would-be observable parameter space. We find that the signal becomes completely inaccessible to Advanced LIGO, independently of the parameters. In other words, there is no solution in the parameter space $\lbrace w_{\rm s}, f_{\rm RD}, H_{\rm inf} \rbrace$ compatible with a detection at LIGO, while not violating at the same time the BBN/CMB constraints [c.f.~Eqs.~(\ref{BBNbound}), (\ref{CMBupperbound})]. This is independent of whether we consider the current run O2 or the projected run O5. In the case of LISA, a small region of parameter space remains still probe-able. This is depicted in the plots of Figure~\ref{fig:remainingLISAregion}, which show clearly that the initially large portion of parameter space compatible with a signal detection by LISA, shrinks into a very small region. This region corresponds to scenarios with a large inflationary scale $H_{\rm inf} \gtrsim 0.1~H_{\rm max}$, transitioning into a RD stage at a low temperature $1~{\rm MeV} \lesssim T_{\rm RD} \lesssim 215$ MeV (i.e.~with the SD epoch spanning for many decades in energy scale, or with equivalently $10^{-11}~{\rm Hz} \lesssim f_{\rm RD} \lesssim 5.5\cdot 10^{-9}~{\rm Hz}$), and where the stiff EoS is confined within a narrow range of small values, $0.48 \lesssim w_{\rm s} \lesssim 0.56$ [c.f. the end of Section \ref{subsec:BBNandCMBbounds}].

We have also studied the impact of adding a small red tilt into the GW spectrum. We find that slightly lower values of $H_{\rm inf}$, higher values for $f_{\rm RD}$, and small values of $w_{\rm s}$, are no longer probe-able by LISA, see left panel of Figure~\ref{figure:tensortiltcorrected}. Overall, the probe-able regions still represent a small island of the parameter space. The inclusion of a constant tilt (even in the maximum case $n_t = -0.008$ allowed by CMB observations) represents therefore only a very small correction. We have considered as well the spectral correction in GW spectrum due to the inclusion of changes in the number of relativistic $dof$ during RD. If the temperature of the radiation fluid component at the time of the SD-to-RD transition is large enough, $T_{\rm RD} > T_{\rm QCD} \simeq 200$ MeV, this does not affect the parameter constraint analysis (as the GW signal does not change within the observable frequency range). However, for low reheating temperatures $T_{\rm RD} \lesssim T_{\rm QCD}$ (as actually required by the surviving probe-able region of parameter space, for which $T_{\rm RD} \lesssim 150$ MeV), this effect yields a correction in the GW spectrum amplitude of a factor $\sim 2$ (in the frequency range of interest). The impact of this in our parameter constraint analysis is that the remaining probe-able regions by LISA are displaced towards larger values of $f_{\rm RD}$, and become slightly wider in the range of $w_{\rm s}$, see the right panel in Figure~\ref{figure:tensortiltcorrected}. The probe-able region now extends down to $w_{\rm s} \simeq 0.46$. Overall, this effect is still a small correction, which does not change the fact that the parameter space region probe-able by LISA still represents a very small island of the full parameter space. We note that neither of these corrections, adding a small tilt or changing the number of $dof$, change the fact that there is no surviving parameter space compatible with having a signal detected by advanced LIGO O2/O5 while not violating the BBN constraint. Finally we note that the expected anisotropic stress due to the presence of the free-streaming relativistic SM neutrinos does not affect the GW signal in the frequency range that is relevant to our analysis.

A stiff epoch is a crucial aspect in various early Universe scenarios, for instance the {\it Quintessential inflation}~\cite{Peebles:1998qn,Peloso:1999dm,Huey:2001ae,Majumdar:2001mm,Dimopoulos:2001ix,Wetterich:2013jsa,Wetterich:2014gaa,Hossain:2014xha,Rubio:2017gty}, where inflation is followed by an epoch dominated by the kinetic energy of the inflaton, with the potential adjusted to describe the observed dark energy today. Various mechanisms for excitation of the radiation component that eventually will dominate the energy budget in these scenarios have been proposed and worked out in detail~\cite{Felder:1999pv,deHaro:2017nui,deHaro:2017bha,Haro:2018zdb,deHaro:2019oki} such that the duration of the SD phase is controlled by the relevant parameters involved in each construction. One of such mechanisms is the original {\it gravitational reheating} model~\cite{Ford:1986sy,Spokoiny:1993kt}, where the Universe is reheated by the decay products of inflationary spectator fields excited during or towards the end of inflation. Basic implementations of this idea have been shown however to be inconsistent with BBN/CMB constraints~\cite{Figueroa:2018twl}, unless unnatural $ad~hoc$ constructions with many fields and identical tuned properties are accepted. Variants of the idea that avoid the previous inconsistency have been proposed, like the {\it Higgs-reheating} scenario~\cite{Figueroa:2016dsc}, where the Standard Model (SM) Higgs is a spectator field with a non-minimal coupling to gravity, and the Universe is reheated into SM relativistic species (decay products of the Higgs) during the SD period. The same mechanism can actually be realized with generic self-interacting scalar fields other than the SM Higgs~\cite{Dimopoulos:2018wfg}, see~\cite{Opferkuch:2019zbd} for a recent re-analysis of the idea. In general, models with blue-tilted inflationary GW spectrum due to the presence of an SD epoch have been studied in various contexts in the past, see e.g.~\cite{Giovannini:1998bp, Giovannini:1999bh, Riazuelo:2000fc,Giovannini:2008zg, Giovannini:2009kg, Tashiro:2003qp, Sahni:2001qp, Boyle:2005se, Boyle:2007zx, Li:2016mmc, Giovannini:2008tm,Caprini:2018mtu}. Unfortunately, the small island of parameter space compatible with a detection by LISA, $H_{\rm inf} \gtrsim 0.1\,H_{\rm max}$, $1~{\rm MeV} \lesssim T_{\rm RD} \lesssim 150$ MeV, and $0.46 \lesssim w \lesssim 0.56$, does not seem particularly appealing/suitable from the model building perspective: these parameters correspond to scenarios where the SD epoch spans for many decades in energy scale, from the end of inflation till the onset of RD. From the point of view of field theoretical constructions it seems actually difficult to obtain an equation of state within the narrow allowed range. For whichever scenario resting upon an early SD period after inflation, BBN sets stringent constraints (recall Figures~\ref{fig:ruledoutbyBBN} and ~\ref{fig:remainingLISAregion}): if $f_{\rm RD} \lesssim 10^{-6}$ Hz (typically corresponding to a long SD period), the only observable window by LISA is the small island of parameter space quantified just above; if $f_{\rm RD} \gtrsim 10^{-6}$ Hz, the GW signal is then simply not observable. Whichever the scenario studied, this simple frequency domain rule must be taken into consideration when assessing the observability of the inflationary GW background distorted because of the presence of a period with stiff equation of state. In light of our analysis, we find it very unlikely that this GW signal will be detected in the future.

Let us note that in our analysis we have asumed Eq.~(\ref{eq:consistencyCondition}) for the spectral tilt of the tensor modes  excited during inflation. This is a relation expected whenever inflation is well described by a quasi-{\rm de Sitter} background, as we assume in this work. In alternative models, $n_t$ could be very different. However, except for very concrete inflationary set-up's, the inflationary tensor spectrum rarely exhibits a large blue tilt for the modes excited during inflation. In our analysis we want to constrain the expansion rate of the Universe after inflation, so we avoid on purpose introducing $n_t$ as a free parameter, as this would represent a degenerate constraint between the post-inflationary expansion rate and the freedom to chose $n_t$ during inflation. We rather focused on the ability of detectors to probe the post-inflationary  expansion rate alone, assuming a natural implementation of inflation. Because of this, we have not considered standard constraints due to pulsar timing arrays (PTA) in our analaysis:  PTA cannot set constraints in our case, as BBN imposes a minimum SD-to-RD frequency of the order of $f_{RD} = f_{\rm BBN} \sim 10^{-11}$ Hz. This means that even in the highest possible slope of the tensor spectrum, when $h^2\Omega_{GW} \propto f$ for $w \simeq 1$, the tensor background can only be at most $h^2\Omega_{GW} = h^2\Omega_{GW}^{(0)}\large{|}_{\rm (plateau)}\times \tilde{\mathcal{A}}_s\times(f_{\rm PTA}/f_{\rm BBN}) \sim 10^{-13}$ at the typical frequencies of PTA experiments $f_{\rm PTA} \sim 10^{-8}$ Hz. Such large tilt are however forbidden due to the BBN constraint, as shown in our analysis, c.f.~Fig.~\ref{fig:remainingLISAregion}. Therefore,  at the PTA frequencies, taking into accout the BBN constraint for the lowest values of $f_{\rm RD}$ , the value of GW background at $f_{\rm PTA}$ is always below any of the projected sensitivities expected for present and future PTA experiments, in particular $h^2\Omega_{GW}(f_{\rm PTA}) \lesssim 10^{-15}$. For constraints on GW signals assuming that $n_t$ can be a free variable, see e.g.~\cite{Zhao:2011bg,Zhao:2013bba,Kuroyanagi:2014nba,Jinno:2014qka,Lentati:2015qwp,Lasky:2015lej,Arzoumanian:2015liz,Liu:2015psa,Kuroyanagi:2018csn,DEramo:2019tit,Bernal:2019lpc}.

As a final remark, let us mention that if we consider an inflationary tensor tilt with running, or a smooth transition from inflation into the SD epoch, there are possible ways to reduce the tension with the BBN/CMB bounds, and hence to enlarge the probe-able parameter space of the GW signal. In realistic inflationary models, typically the deviation from slow-roll becomes more noticeable towards the end of inflation, naturally leading to a running in the tensor tilt. This will reduce further the amplitude of the GW modes in the high frequency end of the spectrum, but this becomes a model dependent computation. For standard single field inflation monomial potentials, one may obtain a reduction factor of the spectral energy amplitude at the highest frequency mode of the order of $\lesssim 0.1$. Furthermore, the transition into an SD era is also a model dependent mechanism, and if the transition were smooth, say lasting for a few Hubble times since the end of inflation, this would also reduce the amplitude at the high frequency modes of the spectrum (as the earliest modes reentering the horizon would not evolve in an SD background yet, or the initial stiffness of the EoS would be softer, i.e.~closer to $1/3$ than to $1$). This question is, again, a highly model dependent one. To assess how our results may change in light of these two effects, specific details regarding these effects need to be worked out for each scenario that one is analyzing.\vspace*{0.5cm}

{\it Acknowledgements}. We thank Benjamin Stefanek for commments on the manuscript. The work of DGF was supported partially by the ERC-AdG-2015 grant 694896 and partially by the Swiss National Science Foundation (SNSF).

\bibliography{referencesSTIFF}
\bibliographystyle{JHEP}

\end{document}